\documentclass[10pt,journal,compsoc]{IEEEtran}
\usepackage{amsmath,amsfonts}
\usepackage{algorithmic}
\usepackage{algorithm}
\usepackage{array}
\usepackage{textcomp}
\usepackage{stfloats}
\usepackage{url}
\usepackage{verbatim}
\usepackage{graphicx}
\usepackage[switch]{lineno}
\usepackage{cite}
\usepackage{hyperref}
\usepackage{booktabs}
\usepackage{paralist}
\usepackage[nameinlink,noabbrev]{cleveref}
\usepackage{listings}
\usepackage[table,dvipsnames]{xcolor}
\usepackage{multirow}

\usepackage{caption}
\usepackage{subcaption}

\newcommand{\expert}[1]{[\textit{id:#1}]}
\newcommand{\ReplicationPkg}{\cite{replicationpackage}}

\newcommand*{\appNoSpace}{\texttt{{SLEEC@run.time}}}
\newcommand*{\app}{\texttt{{SLEEC@run.time~}}}

\newcommand{\circnum}[1]{\textcircled{\scriptsize #1}}

\lstdefinestyle{customSleec}{
    basicstyle=\footnotesize, 
    keywordstyle=\color{NavyBlue}\texttt,
    morekeywords={IF,THEN,UNLESS,IN,WHICH,CASE,WHEN,SCOPE,RULE,OTHERWISE,AFTER,WITHIN,AND,OR},
    morekeywords=[2]{when,then, in, which, case, unless},
    keywordstyle=[2]{\bfseries\itshape},
    escapeinside={(*@}{@*)},
    breaklines=true, 
    breakatwhitespace=true,
    columns=fullflexible, 
    numbersep=8pt,   
}

\lstdefinestyle{ASMmodel}{
  language=AsmetaL,
  frame=lines,
  columns=fullflexible,
  numbers=left,
  numberstyle=\tiny\color{gray},
  numbersep=4pt,
  xleftmargin=12pt,
  framexleftmargin=12pt,
  basicstyle=\scriptsize\sffamily,
  lineskip=1.2pt,
  captionpos=b
}

\lstdefinelanguage{AsmetaL} 
{
morekeywords={module, par, endpar, if, endif, then, else, seq, endseq, signature, definitions, asm, import, function, domain, main, rule, macro, invariant, over, choose, let, endlet, with, ifnone, forall, abstract, default, init, do, agent, dynamic, controlled, monitored, in, out, static, derived, subsetof, switch, case, endswitch, enum, CTLSPEC, ctlspec, LTLSPEC, ltlspec, JUSTICE},
sensitive=true, 
morecomment=[l]{//},
morecomment=[s]{/*}{*/},
commentstyle=\color{gray}\itshape,
literate=
    {<<}{{{$\langle\langle$}}}2
    {>>}{{{$\rangle\rangle$}}}2,
keywordstyle=\color{violet}\bf,
morecomment=[l][\color{white}\tiny]{'},
morestring=[b]",tabsize=1, columns=fullflexible, basicstyle=\scriptsize\sffamily, captionpos=b}

\newcommand{\kwIF}{\textcolor{NavyBlue}{\texttt{IF}}}
\newcommand{\kwTHEN}{\textcolor{NavyBlue}{\texttt{THEN}}}
\newcommand{\kwUNLESS}{\textcolor{NavyBlue}{\texttt{UNLESS}}}
\newcommand{\kwIWC}{\textcolor{NavyBlue}{\texttt{IN WHICH CASE}}}
\newcommand{\kwAFTER}{\textcolor{NavyBlue}{\texttt{AFTER}}}
\newcommand{\kwOTHERWISE}{\textcolor{NavyBlue}{\texttt{OTHERWISE}}}
\newcommand{\kwWITHIN}{\textcolor{NavyBlue}{\texttt{WITHIN}}}
\newcommand{\kwAND}{\textcolor{NavyBlue}{\texttt{AND}}}
\newcommand{\kwOR}{\textcolor{NavyBlue}{\texttt{OR}}}

\newcommand{\base}{\textsc{[base rule]}}
\newcommand{\hc}[1]{\textsc{[hc{#1}]}}

\usepackage{tikz}

\newcommand{\prinplus}[1]{%
  \tikz[baseline=(X.base)] 
  \node[fill=black!10,
        draw=green!60!black,
        rounded corners=2pt,
        inner xsep=3pt,
        inner ysep=1pt] (X)
  {\tiny\textbf{+}\,\textsc{#1}};
}

\newcommand{\prinmin}[1]{%
  \tikz[baseline=(X.base)] 
  \node[fill=black!10,
        draw=red!60!black,
        rounded corners=2pt,
        inner xsep=3pt,
        inner ysep=1pt] (X)
  {\tiny\textbf{-}\,\textsc{#1}};
}

\lstdefinelanguage{EBNF}{
  morecomment=[l]{//},
  morecomment=[s]{/*}{*/},
  morestring=[b]",
  sensitive=true,
  morekeywords={IF, THEN, UNLESS, IN, WHICH, CASE, AFTER, WITHIN, OTHERWISE, SCOPE, RULE, OR, AND},
  morekeywords=[2]{COND_PREDICATE, OBL_ATOM}
}

\lstdefinestyle{ebnf}{
  language=EBNF,
  basicstyle=\footnotesize,
  commentstyle=\itshape\color{gray},
  stringstyle=\color{gray},
  keywordstyle=\color{NavyBlue}\ttfamily,
  keywordstyle=[2]\ttfamily\bfseries,
  showstringspaces=false,
  columns=fullflexible,
  keepspaces=true,
  frame=single,
  rulecolor=\color{black},
  tabsize=2,
}

\begin{document}

\title{Runtime Enforcement for Operationalizing \\ Ethics in Autonomous Systems}

\author{
Martina De Sanctis, Gianluca Filippone, Paola Inverardi, Raffaela Mirandola, Sara Pettinari, Patrizia Scandurra
\thanks{
M.~De Sanctis, G.~Filippone, P.~Inverardi, and S.~Pettinari are with Gran Sasso Science Institute (GSSI), L'Aquila, Italy - e-mail: \{martina.desanctis, gianluca.filippone, paola.inverardi, sara.pettinari\}@gssi.it
}
\thanks{
R.~Mirandola is with Karlsruhe Institute of Technology, Karlsruhe, Germany - e-mail: raffaela.mirandola@kit.edu
}
\thanks{P.~Scandurra is with the University of Bergamo, Bergamo, Italy - e-mail: 
patrizia.scandurra@unibg.it}
}


\makeatletter
\def\ps@IEEEtitlepagestyle{
        \def\@oddfoot{\mycopyrightnotice}
        \def\@evenfoot{}
}
\def\mycopyrightnotice{
        {\footnotesize
                \begin{minipage}{\textwidth}
                        \centering
                        \textcopyright~{\it ``This work has been submitted to the IEEE for possible publication. Copyright may be transferred without notice, after which this version may no longer be accessible.''}
                \end{minipage}
        }
}

\maketitle

\begin{abstract}
This paper addresses the challenge of operationalizing ethics in autonomous systems through runtime enforcement. It first conceptualizes the system’s \textit{ethical space} and outlines a structured \textit{ethics assurance process}. Building on this foundation, it introduces an \textit{enforcement subsystem} that operationalizes ethical rules, specifically social, legal, ethical, empathetic, and cultural (SLEEC) requirements, through the Abstract State Machine (ASM) formalism. The enforcement subsystem is built on the MAPE-K control-loop architecture for monitoring and controlling the system's ethical behavior, and it relies on an ASM-based runtime model of the ethical rules to enforce. This enables the dynamic evaluation, adaptation, and enforcement of \emph{ethical} behavior within a runtime formal model.
The overall approach, named \appNoSpace, is demonstrated on an assistive robot scenario, showcasing how both the robot’s behavior and the governing ethical rules can dynamically adapt to contextual changes. By leveraging a flexible runtime model, \app accommodates changes such as the addition or removal of SLEEC rules, ensuring a robust and evolvable approach to ethical assurance in autonomous systems. The evaluation of \app shows that it effectively ensures the system's adherence to ethical principles with negligible execution time overhead.
\end{abstract}

\begin{IEEEkeywords}
SLEEC rules, Ethics Enforcement, Models@run.time, Autonomous Systems
\end{IEEEkeywords}

\section{Introduction}\label{sec:introduction}

As AI and autonomous systems become more widespread, concerns are growing about the potential harm that could result from their choices and behavior~\cite{Dignum25}. 
In this context, autonomous agents should act transparently in accordance with recognized ethical norms, guidelines, and principles~\cite{UNESCOGuidelines,Radanliev2025ai-ethics}, possibly adapting their behavior ethically to different users and contexts.
Townsend et al.~\cite{Townsend2022sleec} introduced a methodology for eliciting \emph{social, legal, ethical, empathetic, and cultural (SLEEC) rules} in autonomous systems. Building on this, recent studies (e.g.,~\cite{GetirYaman2024sleec-tk,GetirYaman2025,Troquard2024aaai-sleec,FengMYSBAMTBCCC24,kolyakov2025legos}) have proposed methods to translate high-level normative principles into explicit, formalized, and consistent SLEEC rulesets, supporting automated and informed decision-making in autonomous systems.
However, while the existing work focuses primarily on the elicitation, formalization, validation, and verification of SLEEC rules, there is still a lack of contributions addressing their \emph{operationalization} at later stages, such as implementation and testing, that remain largely unexplored~\cite{shahin2022operationalizing}. Specifically, we refer to solutions for translating ethical principles into concrete designs and implementations that can guide the runtime behavior of autonomous systems.

Building on these premises, this paper addresses the operationalization of ethical requirements (in the form of SLEEC rules) through runtime enforcement in autonomous systems. 
To this aim, in our previous work~\cite{SEAMS2026}, we envisioned an ethics assurance process, namely \appNoSpace, for real autonomous systems and their environments, that covers all phases from the elicitation of ethics requirements to and throughout their runtime operation and evolution.
In particular, to enable an autonomous system to satisfy a set of ethics requirements, the process envisages a runtime ethics enforcement approach that steers the system to stay within ethics-respectful regions of its ethics state space. 
SLEEC rules need to be formally specified in terms of an Abstract State Machine (ASM)~\cite{ASMbook}, and then provided as a runtime model to an enforcer software, which leverages the ASMETA runtime simulator~\cite{BombardaBGRS24}. Taking inspiration from~\cite{Falcone2018,BonfantiRS23}, \app features the design and development of an ethics enforcer software, 
as an autonomic manager that wraps around the autonomous system with a MAPE-K (Monitor-Analyse-Plan-Execute over a Knowledge base) control loop architecture~\cite{KephartC03}. This way, SLEEC rules are used to enforce ethical constraints that bound the system’s runtime behavior. The system’s adaptation logic, instead, can be implemented independently, since the proposed approach does not constitute the primary mechanism for adaptation, nor is it intended as a rule-based adaptation solution.

Specifically, this paper makes three main contributions. 
\begin{enumerate}
    \item We present a realistic assistive-care scenario involving a humanoid robot that supports patients in rehabilitation with structured diet and training programs. The validated reference scenario and its related SLEEC ruleset are made available to provide a reusable benchmark for advancing future research in runtime ethics for autonomous systems.
    \item We conceptualize the notion of \textit{ethics state space} and characterize runtime enforcement of ethical principles in autonomous systems as the process of steering the system within \textit{ethics-respectful} regions of such space.
    \item We concretely realize the \app ethics assurance process. In particular, we:
    (i) specify how SLEEC rules can be formally expressed as ASM models;
    (ii) provide an architecture that implements \app by also realizing the \textit{ethics enforcement subsystem} to keep the system within its ethical state space region;
    (iii) implement and deploy \app in the robotic environment of the aforementioned reference scenario. 
\end{enumerate}

The evaluation of \appNoSpace, in both a simulated environment and on a real robot, shows that it effectively ensures adherence to ethical principles with negligible execution overhead. Moreover, the SLEEC rules are easily modifiable and adaptable, making evolutionary steps in the system straightforward to accommodate. 

The rest of the paper is organized as follows: Section~\ref{sec:background} presents the background of this work. Section~\ref{sec:sleec-syntax} provides the SLEEC rules syntax we use in this work, while Section~\ref{sec:example} describes the assistive robot reference scenario. Section~\ref{sec:framework} conceptualizes the system's ethical space and the enforcement of ethical principles, while Section~\ref{sub:methodology} presents the ethics assurance process and the SLEEC rules formal specification. 
The architecture of \app is described in Section~\ref{sec:architecture}. Section~\ref{sec:validation} reports the qualitative analysis of the SLEEC ruleset through the feedback of the experts, the quantitative analysis of the effectiveness and overhead of the \app approach, the proof-of-concept developed on a real robot, and discusses the threats to validity. Discussion and limitations are reported in Section~\ref{sec:discussion}. Related work is presented in Section~\ref{sec:related}, while Section~\ref{sec:conclusion} concludes the work.

\section{Preliminaries}
\label{sec:background}
This section presents the ethical and theoretical foundations that motivate and support our approach.

\subsection{Social, Legal, Ethical, Empathetic, and Cultural (SLEEC) Rules} \label{sec:sleec}

In response to the concerns surrounding the behavior and impact of AI and autonomous systems, experts such as philosophers, lawyers, engineers, and ethicists are investigating new ways to guide and regulate these technologies. One approach to maintaining oversight involves formulating ethical principles as rules shaping how such systems operate and engage in interactions with humans. 
To support the design of systems with this capability, Townsend et al.~\cite{Townsend2022sleec} introduced the concept of \emph{social, legal, ethical, empathetic, and cultural (SLEEC) rules}, and a methodology to elicit actionable constraints on the system behavior from high-level norms and principles.  
SLEEC rules delineate how the system should respond to specific contextual conditions encountered during execution, rather than determining the system’s overarching objectives or behaviors.
Specifically, a SLEEC rule is made up of a single \textit{default rule} possibly followed by one or more \textit{hedge clauses}~(triggered by ``defeating conditions''~\cite{brunero2022reasons,horty2012}). The default rule follows the structure ``\textit{\textbf{when}} \textit{trigger} \textit{\textbf{then}} \textit{response}''. This specifies an event (the trigger) whose occurrence indicates the need to satisfy the constraints defined in the response~\cite{yaman2023specification}.
However, a rule can be defeated or overcome by hedge clauses~\cite{Townsend2022sleec}. 
These are specified by means of the ``\textit{\textbf{unless}} \textit{condition} \textit{\textbf{in which case}} \textit{obligation}'' construct. Each hedge clause specifies a condition in which the original response should be preempted, and there is an obligation to perform an alternate response, thus enabling ethical-oriented reasoning. 
Hedge clauses are meant to support SLEEC principles, including human dignity, autonomy, beneficence, non-maleficence, privacy, and cultural sensitivity. They do so by either prioritizing or limiting these principles depending on the context and promoting the user’s well-being, thereby guiding the autonomous system’s behavior to accommodate these principles.
A concrete example is given in Section~\ref{sec:example}.

A SLEEC rule governing an autonomous system’s response to contextual changes will look like the following: 
\begin{lstlisting} [style=customSleec]
    when (*@$C_0$@*) then (*@$O_0$@*)
    unless (*@$C_1$@*) in which case (*@$O_1$@*) (*@\circnum{1}@*)
    unless (*@$C_2$@*) in which case (*@$O_2$@*)(*@\circnum{2}@*)
\end{lstlisting}

According to the default rule, when the condition $C_{0}$ is verified, the system should fulfill the obligation $O_{0}$. However, two context-dependent hedge clauses are present: \textcircled{\scriptsize1}~ and ~\textcircled{\scriptsize2}.
The order in which hedge clauses are listed specifies their priority. In this work, we follow the evaluation order of linguistically justified rules as proposed in~\cite{Troquard2024aaai-sleec}, according to which hedge clauses are evaluated in a top-down manner, with the last one taking priority over the others. Hence, referring to the previous example, the clause ~\textcircled{\scriptsize1}~ would be evaluated first, followed by the evaluation of clause ~\textcircled{\scriptsize2}.
Assuming that $C_{0}$ and $C_{1}$ are verified (i.e. $C_{0} \wedge C_{1}$ is evaluated to \textit{true}) while $C_{2}$ is evaluated to \textit{false}, then the clause ~\textcircled{\scriptsize1}~ would be activated and $O_{1}$ must be fulfilled. Whereas, if $C_{0} \wedge C_{1} \wedge C_{2}$ is evaluated to \textit{true}, then the clause ~\textcircled{\scriptsize2}~ would be activated and $O_{2}$ must be fulfilled, as there are no further hedge clauses defeating ~\textcircled{\scriptsize2}~ and $C_{2}$ is the last condition that evaluates to \textit{true}. 

\subsection{ASMs and ASMETA tool set}\label{sub:asmeta}
Abstract State Machines~(ASMs)~\cite{ASMbook} are a formal method for modeling computation that uses first-order structures as their states. These structures provide a mathematically precise way to represent the abstract data types that ASMs operate on, extending the concept of Finite State Machines (FSMs) to work with more complex, richly structured states. 

The basic transition rule of an ASM is the \emph{update} rule (basic unit of rules construction) of form: \textit{\textbf{if} condition \textbf{then} Updates}
where \textit{Updates} is a set of simultaneous assignments  $\mathit{f(t_1,\ldots, t_n)} := v$, being $f$ an n-ary dynamic function, $t_i \in \{t_1, \ldots, t_n\}$ terms, and $v$ the value of  $\mathit{f(t_1,\ldots, t_n)}$ in the next state. 
Transition rules have different constructors depending on the update structure they express, e.g., guarded updates (\emph{if-then-else}, \emph{switch-case}), parallel updates (\emph{par}), non-determinism (\emph{choose}), unrestricted synchronous parallelism (\emph{for-all}), etc.

ASMETA~\cite{BombardaBGRS24} is a set of methods and tools based on ASMs to support the specification, validation, and verification of systems' behavior.
ASMETA provides a simulation engine for executing an ASM alongside a target system to monitor, validate, and enforce correctness properties in dynamic environments~\cite{BonfantiRS23}.
Models@run.time\cite{10.1007/s10270-018-00712-x,10.1007/978-3-642-21292-5_7} are used to maintain a formal, up-to-date representation of a software system during its execution. This runtime model 
serves as a foundation for monitoring, analyzing, and adapting the system at runtime, and is implemented by the ASMETA runtime simulator. 

Among runtime assurance techniques, \emph{runtime enforcement} \cite{Falcone2018,BonfantiRS23} extends runtime verification monitors \cite{10.1007/978-3-642-28891-3_37} by actively intervening to correct deviations. Runtime enforcement is crucial for systems where safety and security are paramount, such as medical devices and security-critical control access software (see approaches \cite{10.1145/3126500,BonfantiRS23,RiganelliMM19}, to name a few).
An \emph{enforcement monitor} (or simply \emph{enforcer}) is a component that observes the system's execution and modifies the system's behavior to ensure compliance with a predefined specification (\emph{enforcement model}) of the enforcement strategy.

\section{SLEEC Rules Syntax} \label{sec:sleec-syntax}

In adopting SLEEC rules, we build on the SLEEC language proposed in~\cite{yaman2023specification}, while introducing a lightweight syntax closer to natural language and tailored to the application context.
Our objective is not to redefine the SLEEC formalism, but to preserve its core structure while extending it with two additional elements: an explicit \textit{scope} declaration and a temporal constraint expressed through an \textit{after} construct.
Note that the syntax presented here is not intended as a fully operational grammar, but rather as a descriptive representation that captures the conceptual structure of SLEEC rules as introduced in~\cite{Townsend2022sleec}.

The resulting syntax is shown in Listing~\ref{lst:syntax}.

\begin{lstlisting}[style=ebnf, caption={SLEEC rules syntax.}, label={lst:syntax}]
Rule ::= ( SCOPE ScopeName )?
               RULE RuleID
               IF Condition THEN Obligation
             ( UNLESS Condition IN WHICH CASE Obligation )*

Condition ::= ConditionComposition ( OR ConditionComposition )*

ConditionComposition ::= ConditionItem ( AND ConditionItem )*

ConditionItem ::= "(" Condition ")" | COND_PREDICATE 

Obligation ::= OblItem ( AND OblItem )*

OblItem ::= OBL_ATOM ( AFTER TimeDuration |
                        WITHIN TimeDuration OTHERWISE OBL_ATOM )?
\end{lstlisting}


Each rule is uniquely identified by a rule identifier, ensuring explicit reference within the ruleset.
A rule may optionally be prefixed by a \texttt{SCOPE} declaration. When present, the rule is evaluated only within the specified operational context prescribed by the declared scope; otherwise, it applies globally.
The introduction of scopes enables contextualizing the reasoning in multi-service robotic systems, where different behaviors should be constrained only in different operational modes or scenarios~\cite{CasadeiAAMAS26}. A scope may represent, for instance, a specific application domain (e.g., hospital or restaurant) or a bounded temporal context (e.g., daytime or nighttime operation).
This enables a form of rule modularization for managing complexity within the SLEEC ruleset specification. Each scope tags a coherent subset of rules that apply under a specific context.
As prescribed by Townsend et al.~\cite{Townsend2022sleec}, the body of a rule is composed of a default rule followed by zero or more ordered \texttt{UNLESS} clauses, acting as hedge clauses. 
For the default rule, we adopted the syntax of~\cite{Troquard2024aaai-sleec}, which replaces \texttt{WHEN} from~\cite{Townsend2022sleec} with \texttt{IF}.

Conditions are defined over evaluable predicates (\texttt{\textbf{COND\_PREDICATE}}).
A predicate may consist of a simple expression or of a more complex one involving relational operators (e.g., \texttt{=}, \texttt{!=}, \texttt{<}, \texttt{<=}, \texttt{>}, \texttt{>=}) and boolean connectives. A condition can then be formed either by a single predicate or by combining multiple predicates through boolean operators such as \texttt{AND}, \texttt{OR}, and \texttt{NOT}, possibly using parentheses for grouping.

Obligations are defined over atomic actions (\texttt{\textbf{OBL\_ATOM}}), corresponding to executable robot capabilities. An obligation may consist of a single atomic action or of a conjunction of multiple actions combined with \texttt{AND}, which are interpreted conjunctively (i.e., all must be executed).
Additionally, an atomic obligation may be associated with temporal constraints: \texttt{AFTER} a time duration indicates a delayed obligation execution; \texttt{WITHIN} a time duration \texttt{OTHERWISE} obligation expresses a deadline together with a fallback obligation in case the primary action is not fulfilled within the specified time.

\section{Assistive Robot Reference Scenario}\label{sec:example}

As a reference scenario, we propose an assistive humanoid care robot deployed in a rehabilitation center to support patients diagnosed with diabetes who are undergoing rehabilitation and need to follow a strict diet and physical training program. The scenario has been designed to capture the complexity of a rehabilitation context, integrating multiple dimensions of patient care, including exercise, nutrition, and interaction with healthcare staff. 
The main mission of the robot is to assist patients in adhering to their prescribed routines by providing personalized guidance, monitoring health-related data, and facilitating communication with healthcare staff when needed. 
This includes instructing on exercises and repetitions, ensuring timely meal delivery per dietary plans, and supporting adherence to medical recommendations.
\Cref{tab:robot_capabilities} describes the robot’s main capabilities within this scenario.

\begin{table}[t]
\caption{Robot capabilities in the reference scenario.}
\label{tab:robot_capabilities} \footnotesize
\begin{tabular}{p{0.2\linewidth} p{0.7\linewidth}}
\toprule 
\textbf{Capability} & \textbf{Description} \\
\midrule
\textit{Continuous Data Monitoring} & Each patient is equipped with a smartwatch that monitors key health parameters such as heart rate, activity level, and glucose trends. This data is transmitted exclusively to the robot, which can exploit it to adapt its behavior in real-time, e.g., by adjusting the exercise routine or alerting healthcare personnel in case of anomalies. \\[1mm]

\textit{Exercise Support} & During training sessions, the robot employs its camera and motion recognition capabilities to monitor and count exercise repetitions performed by the patient. It provides real-time verbal feedback and encouragement through text-to-speech and can display exercises via its integrated tablet display. \\[1mm]

\textit{Interaction with Humans} & The robot can interact with both patients and rehabilitation center personnel (e.g., nurse or physician) using voice recognition and text-to-speech. In case of unexpected or potentially risky situations, the robot can request immediate assistance from staff. It also communicates routine notifications to ensure coordination between patients and staff. \\[1mm]

\textit{Dietary Guidance} & At mealtimes, the robot provides timely reminders regarding the patient’s dietary plan and supports adherence to nutritional recommendations by ensuring meals are consumed according to the prescribed program. \\[1mm]

\textit{Patient's Feedback} & During training sessions and mealtimes, the robot collects patient feedback and preferences, e.g.,~via speech recognition or its tablet display.  
\\
\bottomrule
\end{tabular}
\end{table}

To shape how the assistive robot should operate and interact in an ethical-aware manner with patients, we followed the rule elicitation approach proposed by Townsend et al.~\cite{Townsend2022sleec}. 
In line with their approach, we first identified the ethical principles relevant to the application context and then examined how the robot’s capabilities and operational setting give rise to concrete SLEEC concerns. These concerns were subsequently shaped into preliminary rules and iteratively refined, both internally among authors and with external experts' feedback (cf.~\Cref{sec:validation1}).
This resulted in a set of 9 SLEEC rules that constrain the robot’s behavior according to high-level ethical norms and principles.

Each SLEEC rule is mapped to one or more underlying ethical principles (see \Cref{sec:sleec}) and labeled according to the Social, Legal, Ethical, Empathetic, and Cultural dimensions.
Furthermore, the temporal context specifies the scope of the rules, indicating when they apply (e.g., mealtime between 12:30 PM and 1:00 PM) or whether they can occur at any time during the robot’s operation.

The complete set of SLEEC rules identified for the reference scenario, hereafter named \textit{AssistiveCareRobot}, together with associated and motivated principles and SLEEC labels, is available in Table~\ref{tab:sleec-all} (for readability purposes, the \texttt{SCOPE} and the \texttt{RULE} identifier are presented in separate columns).
For the sake of presentation, we illustrate a representative rule in Listing~\ref{lst:sleec-running}, which is used as a running example throughout the paper.

\begin{lstlisting}[style=customSleec,label=lst:sleec-running,caption=Representative SLEEC rule.]
SCOPE Training Time
RULE S2
IF The user is not exercising THEN Show the next exercise AFTER 1 minute
UNLESS The user did fewer exercise repetitions than expected IN WHICH CASE Encourage the user (*@\circnum{1}@*)
UNLESS The user has already been encouraged IN WHICH CASE Get input from the user through a graphical interface (*@\circnum{2}@*)
UNLESS The user has physical issues resulting from the exercises IN WHICH CASE Notify the user that the session is suspended AND Alert the nurse (*@\circnum{3}@*)
\end{lstlisting}

In this rule, associated with the \textit{training time} scope, the default rule specifies that when a user (i.e., the patient) is not exercising, the robot waits for a one-minute rest period before showing the next exercise. However, three hedge clauses are present: ~\textcircled{\scriptsize1}~ if the user performs fewer repetitions than expected, the robot must encourage the user to complete the remaining repetitions; ~\textcircled{\scriptsize2}~ if the user has already been encouraged, the robot uses its graphical interface to ask about the user’s current condition and preference, such as taking a longer rest or drinking some water; ~\textcircled{\scriptsize3}~ if the user experiences any physical problems during the exercises, the robot pauses the session and alerts a nurse, who then checks whether the issue affects the continuation of the session. 
Notably, the proposed sequence of hedge clauses is intentionally designed to ensure that the robot acts proportionally to the situation. Indeed, the physical condition check~(\textcircled{\scriptsize3}) is triggered only after encouragement has already been provided, thereby preventing unnecessary interruptions and allowing the robot to distinguish between temporary lack of motivation and actual physical issues.
However, we acknowledge that situations in which the user exhibits severe physical issues should be addressed within the robot’s default mission logic, i.e., outside the scope of the SLEEC rules.

Each element of this rule is associated with specific ethical principles and SLEEC dimensions. The default rule aligns with the principle of \textit{beneficence}, as it supports physical activity for the user’s well-being. The first hedge clause ~\textcircled{\scriptsize1}~ extends this by further promoting \textit{beneficence} through encouragement aimed at maximizing outcomes. The hedge clause ~\textcircled{\scriptsize2}~ integrates the principles of \textit{autonomy}, through assent and informed consent, and \textit{dignity}, through respectful treatment, while balancing these against \textit{beneficence} by prioritizing respect for the user’s choices over physical activity outcome. Finally, hedge clause ~\textcircled{\scriptsize3}~ represents \textit{dignity} through respectful treatment, \textit{non-maleficence} by minimizing potential harm, and \textit{explainability} by ensuring the user is informed, once again reducing \textit{beneficence} to favor safety and respect. Overall, this rule is labeled under the \textit{Ethical} and \textit{Empathetic} SLEEC dimensions, as it safeguards the user’s well-being while respecting their autonomy.

\begin{table*}[t]
\scriptsize
\centering
\caption{\textit{AssistiveCareRobot} scenario's SLEEC Rules, Principles, and Labels.\\
\scriptsize Description of Principles (``-'' when a principle is violated, ``+'' when a principle is ensured; \\
P: Privacy, D: Dignity, B: Beneficence, NM: Non-Maleficence, A: Autonomy, ET: Explainability\&Transparency).}
\renewcommand{\arraystretch}{1.5}
\begin{tabular}{m{1.5cm} l m{6.8cm} m{6.4cm} m{1.1cm}}
\toprule
\textbf{Scope} & \textbf{Rule} & \textbf{Rule Body} & \textbf{Principles} & \textbf{Labels} \\
\midrule
\cellcolor{green!15}{\textit{Start Training Time}}
&
S1 &
\kwIF\ the user is ready \kwTHEN\ greet in the user's language \kwAND\ start the session \newline
\kwUNLESS\ the user cares about privacy \kwIWC\ greet in the user's language \kwAND\ close the door \kwAND\ start the session \newline
\kwUNLESS\ the room is too warm \kwIWC\ ask for permission to keep the door open \newline
\kwUNLESS\ permission was not asked \kwIWC\ do nothing
&
\base: \prinplus{B} support physical activity; \newline
\hc{1}: \prinplus{P} keeping door closed, \prinplus{B} support physical activity, \prinplus{D} respectful treatment for user preferences; \newline
\hc{2}: \prinplus{NM} minimize harm by refreshing the room, \prinplus{P} respect for user's privacy preferences, \prinplus{A} ask permission to keep door open.
&
Social, Ethical, Empathetic, Cultural \\ \cline{2-5}

\cellcolor{green!15} & 
S1a &
\kwIF\ (permission asked \kwAND\ the user agrees to keep the door open) \kwTHEN\ greet in the user's language \kwAND\ start the session
&
\base: \prinmin{P} violate privacy in favour of non-maleficence, \prinplus{B} support physical activity, \prinplus{A} respect for user consent, \prinplus{NM} minimize harm by refreshing the room.
&
Social, Ethical, Cultural \\ \cline{2-5}

\cellcolor{green!15} &
S1b &
\kwIF\ (permission asked \kwAND\ the room is too warm \kwAND\ the user does not agree to keep the door open) \kwTHEN\ alert the nurse \kwAND\ close the door
&
\base: \prinplus{P} respect for user's privacy preferences, \prinplus{A} respect for user choice, \prinmin{B} physical activity is not performed, \prinplus{NM} training session does not start as the room is too warm and may be dangerous.
&
Ethical \\\hline

\cellcolor{blue!15}{\textit{Training Time}}
& S2 &
\kwIF\ the user is not exercising \kwTHEN\ show the next exercise \kwAFTER\ 1 minute \newline
\kwUNLESS\ the user did fewer repetitions than expected \kwIWC\ encourage the user \newline
\kwUNLESS\ the user has already been encouraged \kwIWC\ get input via the graphical interface \newline
\kwUNLESS\ the user has physical issues resulting from the exercises \kwIWC\ notify user that the session is suspended \kwAND\ alert the nurse
&
\base: \prinplus{B} support physical activity; \newline
\hc{1}: \prinplus{B} maximize outcomes with encouragement; \newline
\hc{2}: \prinplus{A} follow user instructions, \prinplus{D} respectful treatment, \prinmin{B} penalize outcome in favour of dignity and autonomy; \newline
\hc{3}: \prinplus{D} respectful treatment, \prinplus{NM} minimize harm, \prinmin{B} penalize outcome in favour of dignity and non-maleficence, \prinplus{ET} notify user.
&
Ethical, Empathetic \\ \cline{2-5}

\cellcolor{blue!15} & 
S2a &
\kwIF\ (the user is exercising \kwAND\ complains about being tired) \kwTHEN\ encourage the user \newline
\kwUNLESS\ the user expressed the preference to exercise in silence \kwIWC\ do nothing
&
\base: \prinplus{B} support physical activity; \newline
\hc{1}: \prinplus{B} support physical activity, \prinplus{A} respect preferences.
&
Empathetic \\\hline

\cellcolor{orange!15}{\textit{Anytime}}

& S3 &
\kwIF\ a person asks for user training or medical data \kwTHEN\ share the data \newline
\kwUNLESS\ (the user did not grant consent \kwOR\ the person is unauthorized to access that data) \kwIWC\ do not share data \kwAND\ explain why
&
\base: \prinplus{P} safeguard privacy by following user's privacy consent (user), \prinplus{B} support data sharing for medical purposes (user/person); \newline
\hc{1}: \prinplus{P} safeguard privacy (user), \prinplus{ET} explain why data cannot be shared (person). 
&
Legal, Ethical \\ \cline{2-5}

\cellcolor{orange!15} & 
S4 &
\kwIF\ (the user asks for food \kwAND\ it is not meal time) \kwTHEN\ explain why the user cannot eat at the moment \newline
\kwUNLESS\ the user has low glucose level \kwIWC\ give dietary-approved snack \kwAND\ inform the nurse
&
\base: \prinplus{NM} minimize harm by not giving food, \prinplus{ET} explain why the user can not eat at the moment, \prinplus{B} maximize dietary objectives and compliance, \prinmin{D} imposition, \prinmin{A} not granting autonomy; \newline
\hc{1}: \prinplus{NM} minimize the risk of reaching hypoglycemia, \prinplus{A} grant permission to the user, \prinplus{D} accomodating user's will.
&
Ethical, Empathetic \\\hline

\cellcolor{purple!15}{\textit{Mealtime}}
& S5 &
\kwIF\ the user is not yet ready \kwTHEN\ remind the user to eat \newline
\kwUNLESS\ the user is sleeping \kwIWC\ gently wake up the user \kwWITHIN\ 5 minutes \kwOTHERWISE\ alert nurse \newline
\kwUNLESS\ the user is in the REM stage of sleep \kwIWC\ do not wake up the user \kwAND\ inform nurse \newline
\kwUNLESS\ the user is at risk of hypoglycemia \kwIWC\ alert the nurse
&
\base: \prinplus{B} support dietary recommendations; \newline
\hc{1}: \prinplus{B} support dietary recommendation, \prinmin{D} penalize dignity in favour of beneficience; \newline
\hc{2}: \prinplus{D} respectful treatment of not waking up, \prinmin{B} outcome not maximized as user does not eat at the given time, \prinplus{NM} minimize harm of not waking up during the REM stage; \newline
\hc{3}: \prinplus{NM} minimize harm of hypoglycemia.
&
Social, Ethical \\ \cline{2-5}

\cellcolor{purple!15} &
S6 &
\kwIF\ the user is ready \kwTHEN\ deliver meal portions \newline
\kwUNLESS\ the user wants to eat something outside the dietary plan \kwIWC\ explain why the user should adhere to the diet \kwAND\ deliver meal portions \newline
\kwUNLESS\ (the results of the training exercise allow for a different food \kwOR\ the user is particularly distressed) \kwIWC\ deliver dietary-approved different food
&
\base: \prinplus{B} support dietary recommendations; \newline
\hc{1}: \prinplus{NM} minimize harm of not following diet, \prinmin{A} not granting autonomy, \prinmin{D} imposition on what to eat, \prinplus{ET} explains why the user should adhere to the diet; \newline
\hc{2}: \prinplus{B} still following dietary recommendations, \prinplus{A} grant permission to the user, \prinplus{D} user's will is accommodated, \prinplus{NM} delivered food is still approved by the diet.
&
Ethical, Empathetic \\ \bottomrule
\end{tabular}
\label{tab:sleec-all}
\end{table*}

\section{Runtime ethics enforcement}\label{sec:framework}
We provide here concepts and definitions to characterize the ethical aware behavior of an autonomous system. Ethical awareness entails applying ethical reasoning and principles to evaluate alternative courses of action and to make informed decisions. We 
provide
the notion of the \emph{system}, and 
its \emph{ethics state space} relative to the ethical principles of the users it interacts with. 

We assume a system $S = <AutSys, Env, Humans>$ where $AutSys$ is the autonomous system we aim to make ethical-aware, $Env$ is the physical context in which $AutSys$ operates, comprising environmental entities that interact with and influence $AutSys$, $Humans$ are those who may interact with $AutSys$, either directly (e.g., users assisted by the care robot) or indirectly (e.g., individuals sharing the same environment). 

During operation, $AutSys$ perceives parts of $Env$ through its $Probe$ interface (or set of probe events) and reacts accordingly by affecting parts of $Env$ through its $Effector$ interface (or set of effector events).

An ethics state prioritizes ethical behavior in all the system's actions, according to the ethical principles of the $Humans$ with which $AutSys$ interacts.

We consider $\Sigma$ the \emph{system's ethics state space}, describing both $AutSys$ and its $Env$ with respect to the ethical principles of $Humans$, e.g.,~human dignity, beneficence, autonomy~\cite{Townsend2022sleec}.
We classify $\Sigma$ into \textit{ethics-disregarding}, \textit{ethics-critical}, and \textit{ethics-respectful} regions based on how the system's state maps to the set of ethical principles. 

To formalize this mapping, we introduce an \emph{ethical criterion} $\theta$, shaping how an $AutSys$ should operate and engage in interactions with humans.
Therefore, $\theta$ serves as the basis for identifying the \emph{ethics-respectful} region within the state space and guides the system's ethical behavior.
The ethical principles allow for the identification of the subset of the system's state space $\Sigma$ consisting of all those states where the system's behavior satisfies the criterion  $\theta$. We call this subset of states \emph{ethics-respectful region} and denote it by $\theta(\Sigma)$.
In particular, in our approach, the oversight of $AutSys$ is maintained by a set of rules modeling ethical principles, namely the SLEEC rules. In this context, the ethics-respectful region is defined as the situation in which none of the SLEEC rules is triggered, assuming that the ruleset is complete with respect to the ethical principles of $Humans$ for the context in which it was defined.

The \emph{ethics-critical region} $\theta_{crit}(\Sigma)$ includes all those states where the behavior of the system may be ethically unacceptable, but for which a sequence of corrective actions exists. When properly enforced, these corrective actions return the system to an ethical state in $\theta(\Sigma)$.
In our approach, the ethics-critical region corresponds to situations in which one or more conditions in one or more SLEEC rules (including both base rules and hedge clauses) are satisfied. In such cases, the fulfillment of the corresponding obligations is triggered. Obligations are carried out through a sequence of actions whose execution returns the system to the ethics-respectful region. In other words, the system may temporarily enter ethics-critical states, but if it possesses the capabilities to return to an ethics-respectful state, it is ethically enforceable. This property is called \emph{ethics enforceability} as described later.

\begin{figure}[htb]
    \centering
\includegraphics[width=0.9\columnwidth]{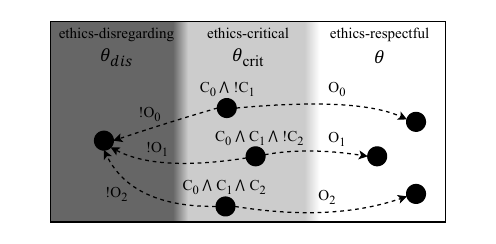} 
      \caption{System's ethics state space and example of the ethics enforcement step considering a generic SLEEC rule.}
\label{fig:state_space}
\end{figure}

The \emph{ethics-disregarding region} $\theta_{dis}(\Sigma)$ includes all states where the system's behavior is ethically unacceptable and from which it is impossible to return to an ethical state.  
In our approach, the system may reach the ethics-disregarding region in three distinct situations:
(1)~the system is not governed by a SLEEC ruleset; in other words, the ruleset is empty;
(2)~the SLEEC ruleset is incomplete, i.e.,~some ethical principles may not be covered by the rules, or some system's functionalities may not be constrained by SLEEC rules, to provide corrective behavioral guidance for $AutSys$;
(3)~$AutSys$ does not fulfill the provided obligations.
For illustration purposes, the ethics state space of a system is illustrated in Figure~\ref{fig:state_space} and described later in this section. 

\smallskip
A system is \emph{ethics enforceable} toward an ethics-respectful state $\theta(\Sigma)$ if, for any system state change, there exists a set of obligations whose fulfillment leads the system back to the ethics-respectful region.
We may achieve ethics enforceability by utilizing a runtime enforcement mechanism. 

\begin{figure}[htb]
    \centering
    \includegraphics[width=0.7\columnwidth]{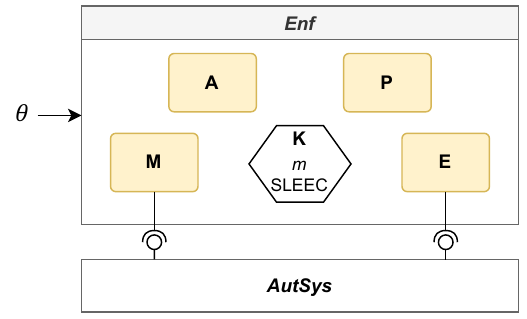}
    \caption{Enforcement by self-adaptation.}
    \label{fig:enforcer}
\end{figure}

The enforcement mechanism can be implemented by monitoring and adaptation. As shown in Figure~\ref{fig:enforcer}, the enforcer $Enf$ manages the target $AutSys$ through a MAPE-K feedback loop. Changes in the ethics space of $AutSys$ are monitored by $Enf$ through probes of the system; $Enf$ can intervene by adapting or modifying the $AutSys$ behavior in response to changes through effectors of the $AutSys$.
Core to the enforcement process is a runtime model $m$ (\emph{enforcement model}) used by $Enf$ for reasoning whether, and which, adjustment actions are required to enforce ethical principles. The runtime model $m$ is part of the knowledge and embeds the \emph{ethics enforcement strategy}. 
Also this strategy can be specified in terms of SLEEC rules.  
Let $C$ represent boolean conditions related to sensed variables in $Probe$, while $O \subset Effector$ represents obligations for the $AutSys$ to act in accordance with ethical principles. $C$ and $O$ correspond to the \emph{conditions} and \emph{obligations} defining SLEEC rules (see the example in section~\ref{sec:sleec}). 
Specifically, the enforcement strategy follows the formalization of SLEEC rules given in \cite{Troquard2024aaai-sleec}, enforcing $O_i$, namely the obligation triggered by the \emph{defeating condition} $\bigwedge\limits_{j=0}^i (C_j) \land \lnot C_{i+1}$.

Figure~\ref{fig:state_space} illustrates the ethics enforcement step over the system's ethics state space performed by the enforcer when the ethics enforcement strategy is formulated in terms of SLEEC rules. In particular, for readability purposes, in Figure~\ref{fig:state_space} we show the pattern of a single SLEEC rule (with three conditions and three obligations) corresponding to that of the generic SLEEC rule illustrated in Section~\ref{sec:sleec}.
For their concrete application, SLEEC rules need to be operationalized. In our approach, they are formalized in terms of ASM transition rules. The resulting ASM model, once validated and verified against errors and conflicts, is in a ready executable form and can be used as a runtime enforcement model.

\begin{figure}[b]
    \centering
    \includegraphics[width=0.7\linewidth]{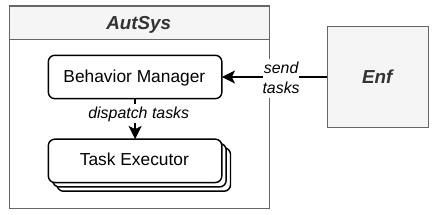}
    \caption{$AutSys$-$Enf$ interaction.}
    \label{fig:interaction}
\end{figure}

It is worth remarking that the ethical space $\Sigma$ does not represent the overall system state. In fact, while the latter captures the system's overall configuration at runtime (comprising, e.g., internal state variables, sensor readings, belief state, mission state), the ethical space represents a view of the system which only considers its ethically-relevant dimensions. That is, $\Sigma$ captures only the state attributes that are relevant to assess the compliance with the ethical principles implied in the interaction between $AutSys$ and $Humans$, whose ethical acceptance is regulated by $\theta$.
Consequently, the enforcement mechanism realized by $Enf$ only operates according to the ethical considerations. Other adaptation mechanisms (e.g.,~optimization, fault recovering, mission replanning), are encapsulated within $AutSys$ and fall outside the scope of this work.
$Enf$ observes ethically-relevant dimensions and, when required, computes obligations to drive the system towards $\theta(\Sigma)$. To this aim, obligations are translated into executable tasks and delivered to the behavior manager of $AutSys$ (see \Cref{fig:interaction}).
The behavior manager is the $AutSys$ component responsible for planning tasks and dispatching them to task executors, which handle their low-level execution. When tasks are received from $Enf$, the behavior manager enforces their execution, overriding other previously-planned ones.
Consequently, although $Enf$ determines whether an obligation must be enforced, its execution is delegated to $AutSys$ relying on the system’s native execution infrastructure. $AutSys$ is also responsible for executing any of the aforementioned adaptation mechanisms, which do not require the involvement of $Enf$. To safely allow a behavioral change (as in the case of executing tasks for enforcing obligations) and, more generally, adaptation, $AutSys$ needs to be in a \textit{quiescent state}~\cite{jeffjeffquiescence,cheng2006model}.
A quiescent state is an operationally stable configuration that safely allows reconfiguration or a behavioral change. In contrast, an ethics-critical state $\theta_{crit}(\Sigma)$ is the state within $\Sigma$ requiring corrective actions according to $\theta$.
While both concepts identify states that precede adaptation, their objectives are different. Quiescent states concern system's operational consistency, whereas ethics-critical states concern ethical considerations. The former ensures that adaptation can be performed safely at the system level; the latter identifies conditions under which obligations can and must be enacted while preserving behavioral coherence with respect to the system's mission and ethical concerns.

\section{\app ethics assurance process}\label{sub:methodology}
This section presents the \app ethics assurance process~\cite{SEAMS2026}, and details its phases. 
It spans from the identification of SLEEC rules to and throughout their operation in the real environment, covering both the \textit{offline} (\texttt{@design.time}) and the online (\texttt{@run.time}) stages.

\begin{figure*}[htb]
    \centering
    \includegraphics[width=1\textwidth]{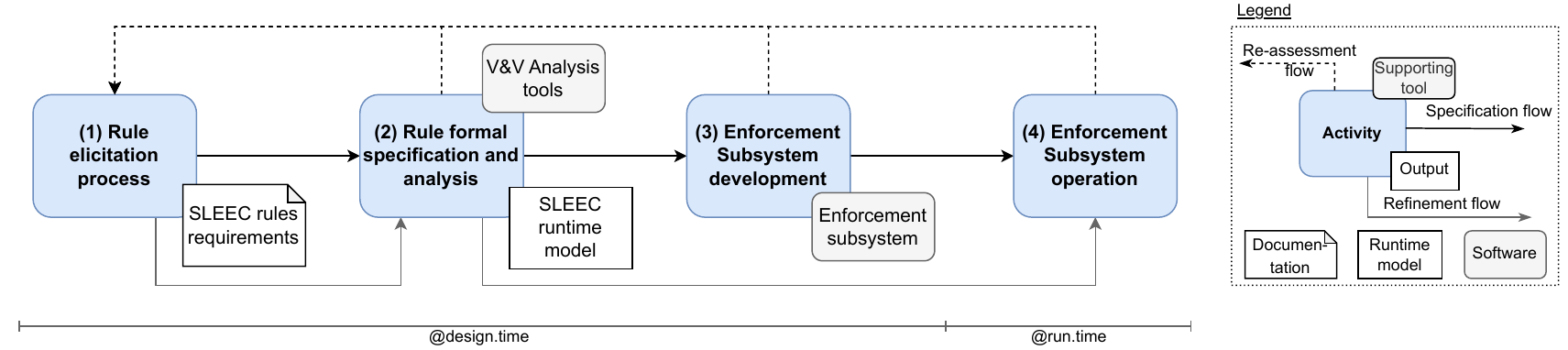}
    \caption{\app ethics assurance process.
    }
\label{fig:process}
\end{figure*}

Figure~\ref{fig:process} illustrates the core lifecycle pillars of the ethics assurance process, from inception to system operation.
The process unfolds sequentially across four main phases:

\begin{enumerate}
    \item \textit{Rule elicitation process}.
    This process follows the iterative approach of \cite{Townsend2022sleec} to derive normative rules aligned with system capabilities and ethical principles. It involves capturing high-level ethical principles, mapping them to system capabilities, and then iteratively identifying relevant SLEEC concerns, resolving conflicts, and refining rules to align with stakeholder values until no new concerns arise. The output of this phase is a complete informal definition of the SLEEC rules that must be enforced by the system.

    \item \textit{Rule formal specification and analysis}.
    The rules elicited in the previous phase are formally specified in ASM (using an automatic model compiler from a SLEEC requirements to the ASMETA language) and analyzed through the ASMETA V\&V techniques (including model animation, validation, verification, and review)  to ensure consistency, correctness, completeness, and minimality. The resulting formally validated and verified \textit{SLEEC runtime model} is already in a final executable form.

    \item \textit{Enforcement Subsystem development}.
    The enforcement subsystem is developed, instantiated, and connected to the target autonomous system, and it implements the $Enf$ component described in \Cref{sec:framework}.

    \item \textit{Enforcement Subsystem operation}.
    The enforcement system is deployed, and the SLEEC runtime model is operationalized. This enables dynamic monitoring system status, making adjustments as enforced by SLEEC rules to ensure regulatory compliance during system execution.
\end{enumerate}

Throughout the lifecycle, stakeholders may revisit and evolve ethical principles or refine the SLEEC rules. This evolution step may be triggered by symptoms or signals emerging from any of the above phases, prompting reassessment and refinement of the normative framework. These updates can inform both the design and runtime phases, ensuring that the system remains aligned with evolving normative expectations. The refinement of SLEEC rules leads to the update of the runtime model, by updating the rules formal specification, which will be uploaded to the running Enforcement Subsystem.

In the following, \Cref{sub:elicitation} and \Cref{sub:formalization} report details for the (1)~rules elicitation processes and (2)~their formal specification and analysis, respectively.
Details on the (3)-(4)~architecture and runtime functionalities of the Enforcement Subsystem are reported in \Cref{sec:architecture}.

\subsection{Rule elicitation process}\label{sub:elicitation}

The rule elicitation process derives the relevant SLEEC rules for the system according to its domain, stakeholders, context, and capabilities. It is realized by following the iterative process defined in~\cite{Townsend2022sleec}, which allows deriving SLEEC rules aligned with system capabilities and ethical values. The process starts with the identification of the high-level ethical norms and principles that are relevant to the application context, drawing from established moral philosophy and normative frameworks (e.g.,~\textit{benevolence}, \textit{non-maleficence}, \textit{autonomy}). These principles are then mapped to the system capabilities, enabling the identification of initial ethical touch points where \textit{base rules} can be formulated. The resulting \textit{base rules} are examined with domain experts and stakeholders to identify additional SLEEC concerns and potential conflicts. Base rules are refined accordingly, by introducing \textit{hedge clauses}, or additional base rules are defined to manage the newly identified concerns or conflicts. This process is iterative, with rules being re-evaluated until all the identified SLEEC concerns are addressed and all the conflicts are resolved. In this way, the SLEEC ruleset obtained through this process is consistent and conflict-free, e.g., there are no multiple rules that require the system to perform conflicting operations at the same time. The process terminates once a stable set of rules has been obtained.

\subsection{Rule formal specification and analysis}\label{sub:formalization}

To correctly capture the operational semantics of a SLEEC rule in ASM, we build on the work in~\cite{Troquard2024aaai-sleec}, which translates SLEEC rules into classical logic. A SLEEC rule can be, therefore, expressed using a nested structure of ASM guarded update rules ({\Cref{sec:sleec-syntax}})
where mutually exclusive guards (based on the triggering conditions, the defeating conditions, and the negation of them) create a workflow that follows only one of the possible execution paths of the ethics state space in \Cref{fig:state_space}. To represent the SLEEC computation pattern, we introduced a new rule constructor \textsf{r\_SLEEC} defined in the textual ASMETA Language (AsmetaL) as follows:
\smallskip
\begin{lstlisting}[language=AsmetaL,basicstyle=\scriptsize\sffamily,lineskip=1.2pt, frame=lines]
//SLEEC rule constructor for 2 hedge clauses
rule r_SLEEC($c0 in Boolean, $o0 in Rule, $c1 in Boolean, $o1 in Rule, 
              $c2 in Boolean, $o2 in Rule) =
	 if $c0 and not $c1 then $o0 
	 else if $c0 and $c1 and not $c2 then $o1 
	 else if $c0 and $c1 and $c2 then $o2 endif endif endif
\end{lstlisting}
where, according to a generic SLEEC rule $r$ expressed as:
\begin{lstlisting}[style=customSleec,mathescape]
    RULE $ID_r$
    IF $C_0$ THEN $O_0$
    UNLESS $C_1$ IN WHICH CASE $O_1$
    UNLESS $C_2$ IN WHICH CASE $O_2$
\end{lstlisting}
the boolean variable \texttt{c0} is the SLEEC's triggering condition $C_0$,  the boolean variables \texttt{c1} and \texttt{c2} are the defeating conditions $C_1$ and $C_2$, the rule \texttt{o0} corresponds to the base rule obligation $O_0$, and the rules \texttt{o1} and \texttt{o2} correspond to obligations $O_1$ and $O_2$.
Such rules serve the purpose to deciding the value of the output function
\(\mathit{outObligation} : \mathit{CapabilityID} \rightarrow \mathit{Boolean}\), where the \textit{True} is assigned to the capabilities of the obligation to enforce on the system. 

\begin{figure}[htb]
\centering
\begin{lstlisting}[style=ASMmodel, deletekeywords={switch, in, case}, columns=fullflexible, numbers=left, caption=Excerpt of a SLEEC model in ASMETAL., label={code:template}]
asm ARIEC
import ../libraries/SLEECLibrary
signature:
//scenario-independent domains
abstract domain Capability
enum domain TCType = {AFTER, WITHIN}
enum domain TimerUnit={NANOSEC, MILLISEC, SEC, MINUTE, HOUR}
...
//scenario-specific domains 
enum domain TimeOfDay = {MEALTIME, STARTTRAININGTIME, ...} 
enum domain CapabilityID = {NOTIFYSESSIONEND,
                            SHOWNEXTEXERCISE,
                            ALERTNURSE, ...}
...
//scenario-independent functions
dynamic out outObligation: CapabilityID -> Boolean
dynamic out outConstraint: CapabilityID
                            -> Prod(TCType,Integer,TimerUnit,CapabilityID)
...
//scenario-specific functions 
dynamic monitored userExercising: Boolean
dynamic monitored fewerExerciseRepetitions: Boolean  
...
//Capabilities:
static notifySessionEnd: Capability
static showNextExercise: Capability
...
static id: Capability -> CapabilityID
...
definitions:
function tooWarm($t in RoomTemperature) = ($t>=26) //Above 26 Celsius
...
rule r_showNextExercise = r_setObligation[showNextExercise,
                                            AFTER,1,MINUTE]
rule r_notifySessionEndAndAlert =
    par r_setObligation[notifySessionEnd] r_setObligation[alertNurse] endpar
...
rule r_Rule2 =
 r_SLEEC[isTrainingTime(timeOfDay) 
            and not userExercising, <<r_showNextExercise>>,
          fewerExerciseRepetitions, <<r_encourage>>,
          userEncouraged, <<r_askUserIntent>>,
          userPhysicalIssues, <<r_notifySessionEndAndAlert>>]
...
invariant inv_1 over outObligation = //never both true
    not (outObligation(id(openDoor)) and outObligation(id(closeDoor)))
...
main rule r_Main =
    seq
        r_Reset[] //reset of out locations
        par r_Rule1[] r_Rule2[] ... r_Rule6[] endpar
    endseq

default init s0: function outObligation($idc in CapabilityID) = undef
\end{lstlisting}
\end{figure}

The \textit{SLEEC runtime model} resulting from the specification of a SLEEC ruleset in AsmetaL is specified in a \texttt{.asm} file structured into five main sections (as shown in Listing~\ref{code:template}):

\begin{compactitem}[-]
  \item \texttt{import}: Imports external definitions, including the \textsf{SLEECLibrary} which provides the SLEEC rule constructor (line 2).
  
  \item \texttt{signature}: Declares domains and functions for representing system knowledge (e.g.,~input events, sensed variables) (lines 3-29). Functions are categorized as \texttt{static} (e.g., the capabilities, lines 25-26) or \texttt{dynamic}, with dynamic functions further classified into \texttt{monitored} (inputs; lines 21-22), \texttt{controlled} (internal state), and \texttt{out} (outputs; lines 16-17).

  \item \texttt{definitions}: Contains static functions, transition rules, and invariants (first-order formulas that must hold in all states, lines 31-47). Invariants are used to express semantic conflicts between obligations. For example, the invariant \texttt{inv\_1} expresses mutual exclusion between the obligations to open and close a door.

  \item \texttt{main rule}: Defines the entry point for computation at each run step (line 48). It may invoke other transition rules by name.
  A run step executes all enabled rules directly or indirectly called from the main rule (SLEEC rules, in our case).
  \item  \texttt{default init}: Initializes dynamic domains and controlled/out functions declared in the \texttt{signature}.
\end{compactitem}

The rule \texttt{r\_Rule2} in Listing~\ref{code:template} (lines 38-43) is an example of a SLEEC rule defined in AsmetaL, corresponding to the example shown in Listing~\ref{lst:sleec-running}.
The conditions and the rules for setting obligations
are passed as arguments to the SLEEC rule. When executed, the conditions of the clauses are evaluated, and the rule of the enabled clause is invoked.
\texttt{r\_notifySessionEndAndAlert} (lines 35-36) is an example of obligation which simultaneously selects (by the \textsf{par} rule) the capabilities \texttt{notifySessionEnd} and \texttt{alertNurse}.
Rule \texttt{r\_setObligation} is a utility rule for setting function \texttt{outObligation(id(\$c)} to true, where \textsf{\$c} is a capability\footnote{\texttt{Capability} is an abstract domain representing the agent's capability objects.
}.

Non-functional requirements can be imposed on obligations. 
As presented in~\Cref{sec:sleec-syntax}, we allow the specification of time constraints for obligations, i.e., response delays \texttt{AFTER} \textit{t time units} (see, e.g.,~rule \texttt{r\_showNextExercise} at line 33), and deadlines \texttt{WITHIN} \textit{t time units} with an optional obligation (\texttt{OTHERWISE}) in case of timeout.
Concerning rule's scope, it is translated in the ASM model by composing the condition defining the scope to the default rule's condition using an \textit{and} boolean operator (see, e.g., rule \texttt{r\_Rule2} at line 39).

To address functional correctness and 
nuanced semantic issues in SLEEC rules, \app relies on the simulation-based techniques available in ASMETA for \emph{model validation}. In particular, we rely on the random simulation mode\footnote{In random mode, values for monitored functions are automatically selected from their codomains.} 
and on \emph{invariant checking}, useful in the early stages of model development to validate the specification against several inputs.
We also use \emph{model review}, a \emph{model verification} technique to check the overall quality of the model by statically capturing
typical modeling errors, including detecting inconsistent function updates\footnote{It is when a function is inconsistently updated to two different values by different, conflicting rules, at the same time. } 
or dead specification parts (rules that are never triggered) due to overspecification. 
The details of such analysis are omitted, as they are beyond the scope of this paper; they can be found in \cite{ScandurraEtAlABZ2026}.

\section{\app architecture}\label{sec:architecture}

\begin{figure}[htb]
    \centering
    \includegraphics[width=\linewidth]{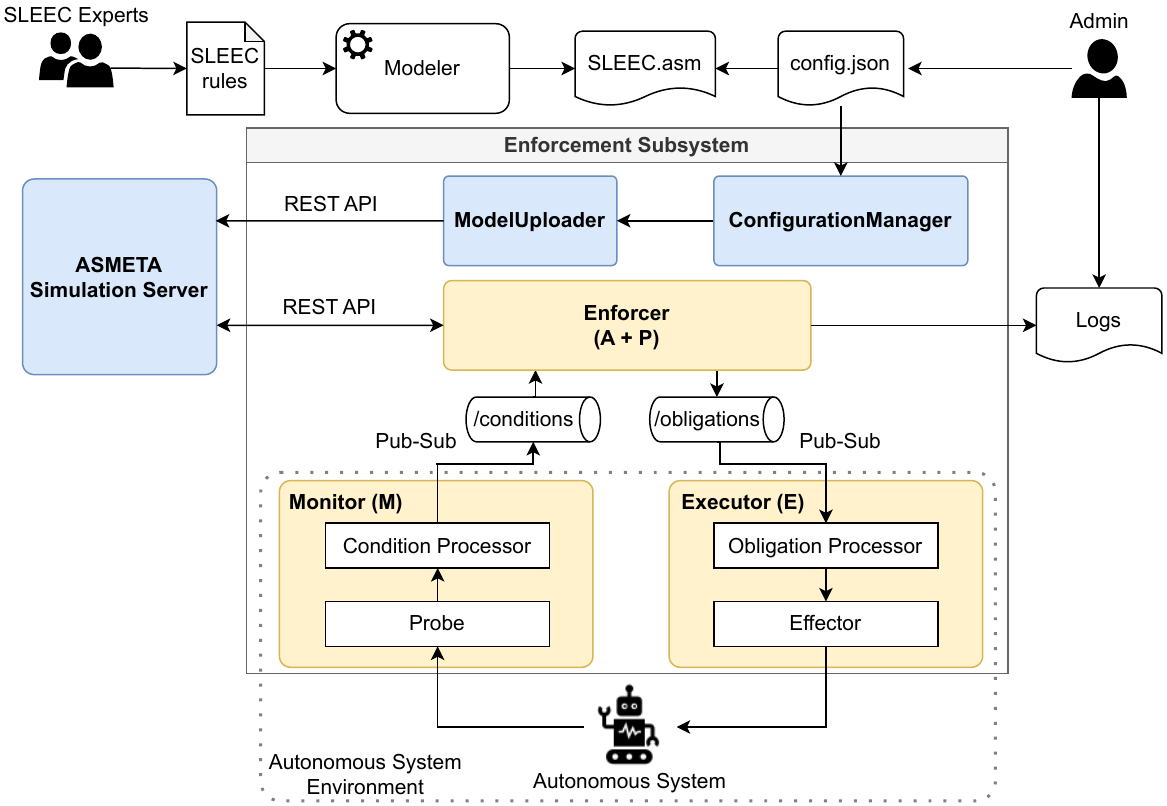}
    \caption{\app architecture.}
    \label{fig:architecture}
    \vspace{-0.2cm}
\end{figure}

Figure \ref{fig:architecture} outlines the architecture of \appNoSpace, which instantiates the \textit{Enforcement by self-adaptation} pattern in \Cref{fig:enforcer} (\Cref{sec:framework}). In \appNoSpace, the target Autonomous System ($AutSys$ in \Cref{sec:framework}) is equipped with a software layer (the \textit{Enforcement Subsystem}) designed to enforce ethics at runtime by implementing the $Enf$ component in \Cref{fig:enforcer}.

The main components of the \textit{Enforcement Subsystem} are the ones that realize the enforcement MAPE-K loop: the \textit{Monitor} (M), the \textit{Enforcer} (A and P), and \textit{Executor} (E).

The \textit{Monitor} component features a \textit{Probe} module that gets the runtime conditions ($Env$ in \Cref{sec:framework}) required for the ethics enforcement by leveraging the autonomous system's perception capabilities, accessed through interfaces offered by the system. The \textit{Condition Processor} module elaborates the low-level sensor data into the higher-abstraction level conditions relevant for the rules evaluation as specified in the SLEEC runtime model, hence raising the $Env$ perception to the SLEEC's abstraction level. On the conditions update, the Monitor triggers the \textit{Enforcer} for the analysis and planning.

The \textit{Enforcer} is in charge of: (i) detecting ethical concerns and triggering conditions based on SLEEC rules (A), and (ii) formulating corrective actions in terms of obligations (P). Both these activities are embedded into the SLEEC runtime model, which is run externally by the \textit{ASMETA Simulation server}. The latter runs the runtime enforcer by exposing RESTful APIs for the upload and execution of the SLEEC runtime model.
The \textit{Enforcer} component is in charge of starting and stopping the SLEEC runtime model execution, and triggering a model execution step through the \texttt{/step} endpoint.
If the system is in the ethics-critical $\theta_{crit}(\Sigma)$ state, the \textit{Enforcer} produces a plan as an output of an execution step.
It consists of a set of obligations (each potentially including parameters such as delays, durations, or fallback obligations in case of timeouts) to be enforced to bring the system to the ethics-respectful $\theta(\Sigma)$ state.
The \textit{Enforcer} also collects system quality metrics and logging data for debugging and manual inspection.

The \textit{Executor} component, through the \textit{Obligation Processor}, translates the received high-level obligations into a sequence of concrete, executable actions (i.e., tasks) for the Autonomous System which realizes them. 
The \textit{Effector} module interacts with the autonomous system through its interfaces for demanding the task execution to the Autonomous System. Importantly, the Effector module does not implement the physical effectors of the Autonomous System (e.g., robot actuators), nor it coordinate the system’s behavior at a low level.
Rather, it is the component of the Enforcement Subsystem (the managing subsystem) that acts on the Autonomous System (the managed subsystem) by invoking the interfaces exposed by the latter. The Autonomous System handles the execution of such actions by exploiting the system's behavior manager for normal mission execution (as described in \Cref{sec:framework}, see \Cref{fig:interaction}).

Both the \textit{Monitor} and \textit{Executor} components require communication with the autonomous system and include dedicated submodules that serve as communication layers. 
Hence, they are implemented to seamlessly integrate with the autonomous system's interfaces, raising the system's abstraction layer to comply with the more abstract specification of ethics rules.

The MAPE knowledge is partially integrated within the \textit{Monitor} and \textit{Executor} components for the mapping between (i) data coming from the probe interface and the conditions defined within the SLEEC runtime model, and (ii) obligations and functional abilities of the system, respectively. The current conditions and the obligations to enforce, instead, are shared among the components via a Pub/Sub mechanism.

Within \appNoSpace,
\textit{SLEEC Experts} define (and refine) SLEEC rules (requirements documents), which are then translated into SLEEC runtime models, according to phases (1) and (2) in~\Cref{sub:methodology}. The \textit{Admin} is responsible for setting configuration parameters into the configuration file \textit{config.json}, and of observing the overall system by inspecting the produced logs. The \textit{Configuration Manager} component initializes the enforcement framework according to the configuration, then invokes the \textit{ModelUploader} for uploading the required SLEEC runtime model into the ASMETA Simulation Server through its \textit{/upload-model} RESTful endpoints. This structure allows the SLEEC runtime model to be updated within the ASMETA server during the system runtime, being transparent to the \textit{Enforcer Subsystem}, hence allowing the model's re-assessment and refinement, as mentioned in the assurance process in~\Cref{sub:methodology}.

Taking inspiration from the component framework in~\cite{BonfantiRS23}, we developed the enforcer adopting a MAPE-K control loop architecture as shown in Figure~\ref{fig:architecture}. Components in yellow are the constituent elements of the MAPE-K loop, while blue components provide model upload/execution. 

\section{\app Evaluation}\label{sec:validation}

This section presents the \app evaluation results. Section~\ref{sec:validation1} covers expert validation of the \textit{AssistiveCareRobot} scenario’s SLEEC ruleset, assessing alignment with the execution context’s ethical principles. Section~\ref{sec:validation2} provides a quantitative analysis of \appNoSpace’s effectiveness in enforcing ethical behavior and the time overhead of the enforcement subsystem. Section~\ref{sub:proof-of-concept} describes the proof-of-concept developed using a real robot.

\subsection{Qualitative analysis}\label{sec:validation1}

To validate the elicited SLEEC rules, we conducted semi-structured interviews with domain experts, resulting in a reusable ruleset tailored to our scenario and validated through their feedback.
The validation process focused on the following validation question~(VQ):
\begin{enumerate}
    \item[\textbf{VQ}] Do the elicited SLEEC ruleset, their connected principles, and associated SLEEC labels provide a clear, complete, correct, and value-aligned representation of ethical considerations with respect to the reference scenario?
\end{enumerate}
To assess \textbf{VQ}, we engaged academic experts to evaluate the \textit{understandability}, \textit{completeness}, \textit{correctness}, and \textit{alignment} of the SLEEC rules with ethical concerns that may arise in the \textit{AssistiveCareRobot} reference scenario (see Section~\ref{sec:example}). 

\smallskip
\noindent\textbf{Methodology and domain experts selection}. We followed the \textit{qualitative survey empirical standard} and its essential attributes~\cite{acmsigsoft} to perform semi-structured interviews. 
Participants were selected through \textit{convenience sampling}~\cite{stratton2021population}, to gather expert feedback. 
The criteria used for recruiting them were based on their knowledge and expertise with ethical-based systems and reasoning, and SLEEC rules.
For anonymity purposes, the experts are referred to throughout the text as \textit{[id:1]}--\textit{[id:4]}. Each of them has contributed research publications focusing on SLEEC-based approaches. Specifically, they include researchers and faculty members working in computer science-related areas.
Their combined expertise spans logic-based methods, formal verification, and software engineering, while also addressing the design, verification, and ethical implications of autonomous systems.
We eventually interviewed four experts.

Two of the authors conducted synchronous, one-on-one conversations with individual experts, where each conversation was driven by an interview protocol. 
Experts' answers were used to iteratively adjust the ruleset, focusing on the relevance and coherence of answers in relation to the scenario. 
Suggested changes were incorporated once all authors agreed and any issues of clarity or practical applicability were addressed.
The final SLEEC ruleset was used in the implementation and is available online in the replication package~\ReplicationPkg.
Supplementary material also includes the interview guide, details on the interview process, the transcripts of all four interviews, and the complete evolution flow of the SLEEC ruleset.

\smallskip
\noindent\textbf{Results of the validation with domain experts}.
We submitted to the expert's evaluation the SLEEC rules as reported in the ``SLEEC iterations'' document (see \ReplicationPkg{}). The questions in the interview were designed to gather their opinions on the understandability, completeness, correctness, and alignment of the proposed rules. Below is the set of open-ended questions the domain experts were asked. The first three questions (\textbf{Q1--Q3}) were applied to each individual SLEEC rule, while \textbf{Q4} was addressed after reviewing all rules.
\begin{enumerate}
    \item[\textbf{Q1}] Is the SLEEC rule correct and effective in preventing or reducing the system’s unethical behavior?
    \item[\textbf{Q2}] Are the principles correctly mapped to the SLEEC rule, the base rule, and the hedge clauses? Are there any principles that are not covered by this mapping?
    \item[\textbf{Q3}] Is the proposed rule correctly labeled as Social, Legal, Ethical, Empathetic, and Cultural?
    \item[\textbf{Q4}] Are there any SLEEC rules you consider missing?
\end{enumerate}

With respect to \textbf{Q1}, the experts generally rated the SLEEC rules as correct and effective, suggesting only minor textual clarifications to prevent misunderstandings. Additionally, \expert{1} suggested revising the last hedge clause of the first SLEEC in our ruleset (see \ReplicationPkg), later refined in the third iteration by \expert{3}.
SLEEC 1 regulates the robot’s behavior at the beginning of a training session, ensuring appropriate greetings and respect for privacy preferences. In the initial version, the last hedge clause addressed cases of user physical discomfort. 
The refinement aligned the rule with its intent by considering context, i.e., when the user prefers to keep the door closed for privacy but the room is too warm, the robot asks the user’s preference before acting.

Concerning \textbf{Q2}, the experts generally agreed that the principles were correctly mapped, though some missing ones were identified and subsequently added.
Notably, SLEEC 3, which handles requests for a user’s data, was particularly discussed.
While the default rule allows data sharing, its hedge clause prevents it when consent is missing or the requester is unauthorized. Initially, this clause was linked to \textit{autonomy}, as it depends on user consent. 
However, \expert{1} argued that it reflects a legal requirement rather than promoting autonomy, whereas \expert{2} and \expert{4} viewed it as autonomy-preserving.
Following these observations, the mapping was revised according to \expert{1}.
Additionally, \expert{3} noted that since two stakeholders are involved, the principles may apply to one or both of them. This suggestion was adopted by explicitly annotating to which stakeholder each principle applies.

Regarding \textbf{Q3}, experts helped identify and clarify the appropriate labels for each rule, pointing out cases where initial labels were missing or not properly used. Their suggestions were incorporated, resulting in a revised correct labeling. 

Finally, according to \textbf{Q4}, the experts supported the elicitation of SLEEC rules and their association with principles and labels, integrating them into the existing ruleset. Specifically, \expert{1} contributed to the elicitation of SLEEC 2a, which addresses cases when the user complains during exercise, connecting to the discussion in SLEEC 2 (see~\Cref{sec:example}). This new SLEEC rule was further refined by \expert{3}. Additionally, \expert{3} proposed a refinement to SLEEC 1 to better support user autonomy by leveraging two complementary SLEEC rules. This proposal was implemented and lightly adjusted by \expert{4} to avoid inconsistencies.

Concerning \textbf{VQ}, results show that domain experts found the SLEEC ruleset generally clear, correct, and aligned with ethical considerations, though some aspects required further clarification and adjustments to ensure completeness and correctness. 
The iterative structure of the interviews continuously refined the ruleset, with the final expert suggesting only minor adjustments.

\subsection{Quantitative analysis}\label{sec:validation2}
To evaluate the effectiveness of \app in enforcing the system's ethically compliant behavior and the overhead introduced by the Enforcement Subsystem, we implemented the \app architecture (shown in \Cref{fig:architecture}) and conducted experiments by deploying the system in a ROS2-based environment and running it within our reference scenario.
The evaluation was focused on the following evaluation questions (EQs):

\begin{itemize}
    \item[\textbf{EQ1}] How effective is \app
    at enforcing ethical principles, defined as SLEEC rules, at runtime? 
    \item[\textbf{EQ2}] What is the impact of the Enforcement Subsystem in terms of computation overhead? 
    \item[\textbf{EQ3}] How does the overhead of the Enforcement Subsystem scale as the number and size of SLEEC rules increase?
\end{itemize}

By answering \textbf{EQ1}, we aim at assessing the effectiveness of \app in enforcing the defined SLEEC rules, in such a way that it stays in the ethics-respectful space. 
By answering \textbf{EQ2}, we aim at evaluating whether the enforcement process run by the Enforcement Subsystem imposes an acceptable computational overhead in evaluating the runtime context, the SLEEC rules, and enforcing the autonomous system's compliant behavior. 
By answering \textbf{EQ3}, we aim at evaluating how the computational overhead introduced by the Enforcement Subsystem and its components grows when the number and size of the defined SLEEC rules grow, stressing the scalability of \appNoSpace's runtime components when the SLEEC rulesets size goes largely beyond its presumably limited real-world size.
The source code of the implemented system, the collected raw data, and the obtained results are publicly available in the replication package~\ReplicationPkg.

\subsubsection{Experiment setting} 
To address the research questions, we implemented the \app components (\Cref{fig:architecture}) and ran the system for two different experiments.
In the scope of EQ1 and EQ2, we tested \app over the \textit{AssistiveCareRobot} reference scenario. This experiment is realized by providing the ASMETA server with the ASM model encoding the SLEEC rules in \Cref{tab:sleec-all} and generating a set of 750 different test cases, each featuring randomly assigned values to the set of conditions defined in the SLEEC ruleset and monitored by the \textit{Monitor} component. Each of the test cases also reports, as a ground truth, the list of obligations that are expected to be enforced according to the test case's conditions and the SLEEC semantics.
In the scope of EQ3, we generated multiple synthetic ASM models featuring an increasing number of $r$ rules and $c$ clauses (we consider a clause both the base rule and the hedge clauses) for each rule in the model. We generated 110 ASM models, one for each value of $r$ in $\{10, 15, 20, 25, 30, 35, 40, 45, 50, 55, 60\}$ and $c$ in $\{2, 4, 6, 8, 10, 12, 14, 16, 18, 20\}$, and 50 test cases for each model. A monitored condition is generated in the SLEEC ASM model for each of the $r \times c$ clauses, while each test case features a randomly assigned value to such conditions. In total, 5500 test cases were generated and run for testing \app scalability.

\begin{figure}[htb!]
    \centering
    \includegraphics[width=\linewidth]{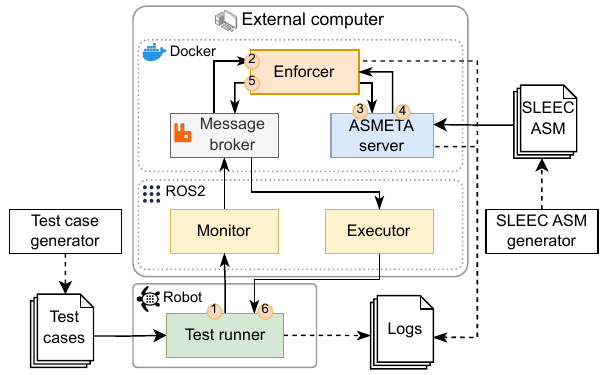}
    \caption{Experimental setting overview}
    \label{fig:experimenal_setting}
\end{figure}

\Cref{fig:experimenal_setting} shows the experimental setting. Before running the experiments, a \textit{Test case generator} generates the test cases for both the tests over \textit{AssistiveCareRobot} and for the scalability tests, while a \textit{SLEEC ASM generator} generates the set of synthetic ASM models used for the scalability tests.
For the experiment execution, the SLEEC ASM model of the experiment (i.e., the one representing the reference scenario, or one of the generated ones for scalability tests) is uploaded to the ASMETA server, while an ad-hoc realized \textit{Test runner} reads each of the test cases defined for the current model, and simulates, for each test case, the prescribed runtime conditions by mocking the robot perception capabilities. The \textit{Monitor} component monitors the simulated conditions and triggers the \textit{Enforcer} component to get the obligation to be enforced. The \textit{Executor} component receives the obligation and translates it into the corresponding sequence of tasks that the system must execute. The test runner, mocking the robot's capabilities and actuators, finally logs the enforced robot behavior for the current test case.

To emulate a real-world deployment, we implemented the Test runner as a ROS node and deployed it on a Turtlebot 4\footnote{\url{https://clearpathrobotics.com/turtlebot- 4/}} robot, equipped with a Raspberry Pi 4B running Ubuntu 22.04 and ROS 2 Humble. We used a Turtlebot robot instead of a real assistive robot for the following reasons: (i) to prevent the simulated conditions from being influenced by the robot’s own perception systems, and (ii) to highlight the generality of our approach by using a platform with limited computational resources, by creating a ``worst-case'' scenario.
The \textit{Enforcement Subsystem} was deployed on an external computer, a Dell Precision workstation equipped with an Intel Xeon W5 and running Ubuntu 22.04. In particular, the \textit{Monitor} and \textit{Executor} components were deployed as ROS nodes running on ROS 2 Humble, 
while the \textit{Enforcer} component and the \textit{ASMETA server} as two separate Docker containers. Additionally, a further Docker container with a RabbitMQ message broker was deployed using the RabbitMQ Docker official image to allow the message passing between the \textit{Monitor}, the \textit{Enforcer}, and the \textit{Executor} through the \texttt{/conditions} and \texttt{/obligations} pub-sub channels. 
Both the robot and the computer were connected to a local WiFi network providing, on average, 3.4 ms round-trip time, measured by sending periodic \textit{ping} messages during the experiment execution.

\subsubsection{Metrics}
While running the experiments, we collected logs from the \textit{Test runner}, \textit{Enforcer}, and \textit{ASMETA server} components.
We also logged the expected behavior for the current test case.
Out of the collected logs, we computed:
(i)~the \textit{Enforcement Subsystem} overhead, as the total time measured by the Test runner from the simulation of a contextual condition to the receiving of the obligation to enforce (steps~\circnum{1}-\circnum{6} in \Cref{fig:experimenal_setting});
(ii)~the \textit{Enforcer} overhead, as the time required by the Enforcer component to perform the analysis and the planning phases of the MAPE-K loop within \appNoSpace, including the time required by \textit{ASMETA server} for running the model enforcement, computed as the time elapsed from receiving the new conditions to sending back the obligation to be enforced (steps~\circnum{2}-\circnum{5}); (iii)~the \textit{ASMETA server} overhead, as the time needed by the ASMETA server for running the model enforcement, computed as the time elapsed from receiving a request via the REST interface to completing the enforcement step and sending the response (steps~\circnum{3}-\circnum{4}).
Moreover, we computed the total number of test cases for which the tasks sequence to be executed by the \textit{Executor}, corresponding to the obligation to be enforced, is aligned with the one prescribed by the ground truth of the test case.

\subsubsection{Results}
Regarding \textbf{EQ1}, the collected data show that the computed obligation to be enforced was consistent with the expected one across all 750 executed test cases. 
This result demonstrates that the SLEEC rules have been correctly formalized as ASM transition rules, and the \textit{Enforcement Subsystem} correctly executes the SLEEC ruleset, consistently with the formalization given in~\cite{Troquard2024aaai-sleec}, by always bringing the system from the ethics-critical $\theta_{crit}(\Sigma)$ region to the ethics-respectful $\theta(\Sigma)$ state space.
This shows the effectiveness of \app at enforcing ethical principles defined as SLEEC rules, at runtime. 

\begin{figure}[htb]
    \centering
    \includegraphics[width=0.8\linewidth]{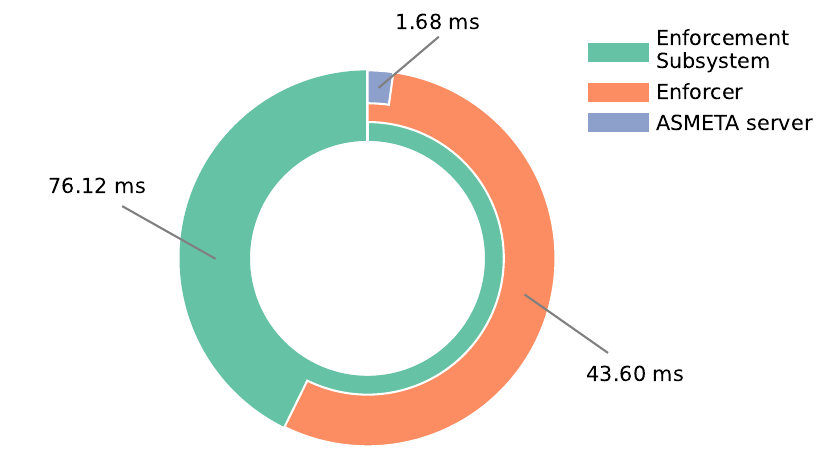}
    \caption{Overhead results}
    \label{fig:overhead_pie_chart}
\end{figure}

Concerning \textbf{EQ2}, we observed that the overall overhead measured, on average, is 76.1 ms (104 ms max., 63 ms min., 5.7 std. dev.). \Cref{fig:overhead_pie_chart} shows the three measured overhead times. Interestingly, around half of the overhead (57\%, 43.6 ms) is due to the Enforcer's processing time. 
This overhead is not affected by network delays between the robot and the computer, as both the Enforcer and the ASMETA server run locally; it likely stems from updating the Enforcer’s knowledge, setting up the REST call across Docker containers, and the ASMETA server’s response time.
The ASMETA server’s overhead averaged 1.68 ms, indicating that enforcing the SLEEC ASM model has minimal impact on the overall process. This efficiency likely stems from the underlying implementation technology.
In fact, while the ASMETA server is realized as a Java application, compiled into an executable .jar file, the Enforcer is realized in Python, which is generally less efficient than Java~\cite{zhang2022quantifying,benchmarksGame}.

\begin{table}[htb]
\centering
\caption{Overhead times overview (data in ms)}
\label{tab:overhead_report} \footnotesize
\begin{tabular}{p{0.32\linewidth} p{0.06\linewidth} p{0.06\linewidth} p{0.06\linewidth} p{0.06\linewidth} p{0.06\linewidth} p{0.06\linewidth}}
\toprule 
\textbf{Component} & \textbf{Avg} & \textbf{Min} & \textbf{Max} & \textbf{75\%} & \textbf{99\%} & \textbf{$\sigma_{std}$}\\
\midrule
\textbf{ASMETA server} & 1.68 & 1 & 64 & 2 & 8 & 1.7 \\
\textbf{Enforcer} & 43.6 & 35 & 297 & 45 & 56 & 5.6 \\
\textbf{Enforcement Subsystem} & 76.1 & 63 & 104 & 79 & 93 & 5.7 \\
\bottomrule
\end{tabular}
\end{table}

\Cref{tab:overhead_report} summarizes all the measured overheads.
It is worth noting that the Enforcer component shows a higher maximum overhead than the overall system. This occurs because, in most cases, updating a single condition triggers the Enforcer’s analysis phase without producing an obligation to enforce. Consequently, this higher internal overhead does not increase the overall overhead, as no obligation is returned to the executor or, in turn, to the robot. Therefore, the robot experiences no delay in executing its tasks.
Moreover, from a statistical point of view, we note that both at the 75th and the 99th percentile, the measured enforcer overhead is very close to both the minimum and the mean values, indicating that the majority of measurements are tightly concentrated around the median. 
The extremely high classical skewness~\cite{Doane01072011} ($\approx$ 20.7) reflects the influence of a few high outliers, while the low Bowley skewness~\cite{weisstein_bowley_skewness} ($\approx$ 0.11), confirms that most of the data lie around the median value, with only very few extreme (higher) values.
It is worth to note that most of higher values occur in the first execution steps of the ASMETA model, likely due to a ramp-up of the Java-based ASMETA server.

In summary, in response to \textbf{EQ2}, we observe that the overall overhead was limited across all 750 test cases. Most enforced actions were received by the robot in less than 0.1 seconds (93 ms at the 99th percentile), indicating good efficiency of the implemented \app approach.

\begin{figure}[htb]
    \centering
    \includegraphics[width=\linewidth]{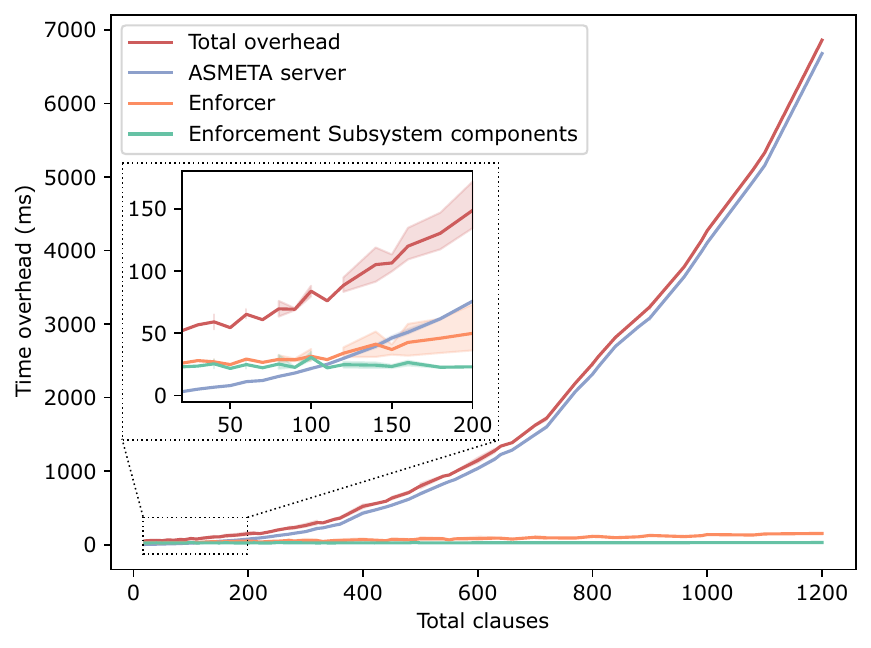}
    \caption{Time overhead per component over SLEEC ruleset size}
    \label{fig:scalability_overview}
\end{figure}

Concerning \textbf{EQ3}, we observed an increasing trend of the measured overhead when the total number of clauses in the generated SLEEC ASM models increases. \Cref{fig:scalability_overview} shows the trend of the total overhead (i.e., the one measured by the Test Runner, representing the steps ~\circnum{1}-\circnum{6}~ in \Cref{fig:experimenal_setting}) and the contribution to the overhead by each component: \textit{ASMETA server}, the \textit{Enforcer} (excluding the ASMETA server, corresponding to steps ~\circnum{2}-\circnum{3}~ + ~\circnum{4}-\circnum{5}), and the \textit{Enforcement Subsystem}'s Monitor, Executor and Message Broker components (corresponding to steps ~\circnum{1}-\circnum{2}~ + ~\circnum{5}-\circnum{6}). Interestingly, for the Enforcement Subsystem's components and for the Enforcer the overhead has minor/negligible increases (the average overhead with 1200 total clauses for the Enforcer is 153 ms). In contrast, the ASMETA server resulted to be the most affected by the increase in the number of clauses within the SLEEC ASM model. When the number of clauses is contained ($\approx$100), the ASMETA server's overhead is lower than the other components, while it outgrows them with bigger ASM models.

Regression analysis revealed that the trend of the ASMETA server overhead with respect to the total number of clauses achieved a strong fit under a log–log model ($R^2 = 0.978$), with estimated exponent $\alpha = 2.07$ suggesting a near-quadratic polynomial growth. In contrast, both linear and exponential models obtained weaker fits ($R^2 = 0.822$ and $0.907$, respectively), further supporting the polynomial trend. Regression with a second-grade polynomial model obtained an accurate fit ($R^2=0.996$), hence confirming this interpretation.
Overall, these results indicate that the scalability of the Enforcer Subsystem is dominated by an approximately quadratic growth in the total number of clauses in the SLEEC ASM model.
This is arguably due to the higher computational complexity of the ASM model enforcement process than the operations performed by the Enforcer and the other Enforcement Subsystem components. They do not consider the whole SLEEC rules, but only the updated conditions and enforced obligations, hence being only partially affected by the growing size of the ruleset. However, while this results in bigger messages to be exchanged, or bigger data structures to be maintained, it does not significantly affect their computation (which mainly consists on checking if a condition has changed and, if so, triggering a model execution step).

\begin{figure}[htb]
    \centering
    \includegraphics[width=\linewidth]{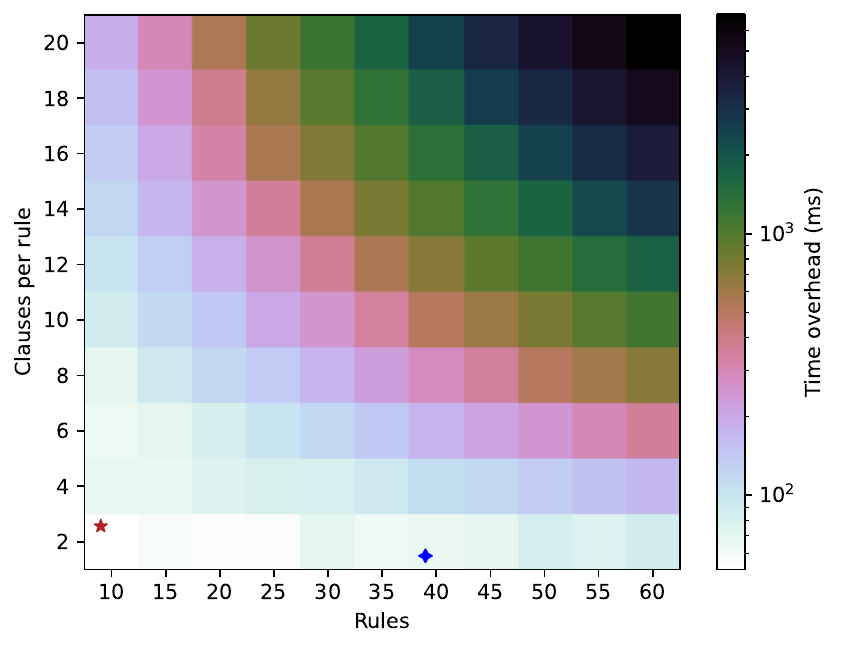}
    \caption{Comparison of overhead in tests run and SOTA SLEEC rulesets dimension}
    \label{fig:sota_heatmap}
\end{figure}

\Cref{fig:sota_heatmap} reports the overhead measured for the ASMETA server according to the number of rules and clauses per rule (i.e., the base rule plus the hedge clauses) in the SLEEC ASM model.
For around half of the combinations, the measured overhead falls below 300 ms, spanning from models with 15 rules having 20 clauses per rule, to models with 60 rules having 4-6 clauses per rule (around 300 clauses in total).
The figure also marks the dimensions of the SLEEC ruleset of the \textit{AssistiveCareRobot} reference scenario (\Cref{sec:example}) and \textit{Autocar}, i.e., the largest use case from~\cite{FengMYBAMTSSIR024} in terms of number of rules\footnote{For comparison purposes, the number of rules and clauses has been normalized by jointly counting the rules, and related clauses, that share the same conditions but different obligations. In our ruleset representation, multiple obligations may be expressed conjunctively within a single clause using the \texttt{AND} operator, whereas in~\cite{FengMYBAMTSSIR024} each obligation would typically be encoded as a separate rule sharing the same triggering condition.}.
They are reported as a red star and a blue diamond, respectively.
The overhead measured on the ASMETA server during our experiments with ASM models of size comparable to the \textit{Autocar} use case averaged 15 ms%
\footnote{\textit{Autocar} features 39 rules and 1.49 clauses per rule; the considered ASM model for this reference features 40 rules and 2 clauses per rule.}, while for \textit{AssistiveCareRobot} averaged 3 ms%
\footnote{\textit{AssistiveCareRobot} features 9 rules and 2.56 clauses per rule; the considered ASM model for this reference features 10 rules and 2 clauses per rule.},
hence supporting the negligibility of the overhead for enforcing ASM models modeling real-world-sized SLEEC rulesets.
As a reference, the ASMETA server overhead measured for \textit{AssistiveCareRobot} in the scope of EQ2, reported in \Cref{tab:overhead_report}, was lower (1.68 ms on average). This difference can be explained by observing that, for EQ2, 750 test cases were run, while for EQ3, only 50 test cases per ASM model were run. This increased the weight, on the average, of the high outliers (as discussed for EQ2, mainly occurring in the first test case run).

In summary, in response to \textbf{EQ3}, we argue that the scalability of \appNoSpace's Enforcement Subsystem is dominated by the ASMETA server quadratic growth over the total number of clauses in the SLEEC ruleset, while the other components are only very marginally affected by the size of the ruleset. However, SLEEC rulesets in the literature~\cite{FengMYBAMTSSIR024} can still be managed without significant impacts on the overall system's performances since their size is limited, while higher overheads are observed with significantly larger rulesets, which are arguably unlikely to be defined in real-world scenarios.

\subsection{Proof-of-concept} \label{sub:proof-of-concept}

To demonstrate the feasibility of \appNoSpace, we developed a proof-of-concept using the ARI humanoid robot\footnote{\url{https://pal-robotics.com/robot/ari/}} from PAL robotics.
The goal was to operationalize the SLEEC rules from the \textit{AssistiveCareRobot} scenario and observe how the robot’s behavior dynamically adapts to remain ethics-respectful.
To this end, we leveraged the facilities provided by the ARI platform to combine some of its native capabilities and abstract them into presentation modules, each representing an obligation (e.g., greeting the user or displaying an exercise).
These modules are triggered through ROS topics, allowing external control by the Enforcement Subsystem through its Executor component.

\begin{figure}[htb!]
    \centering
    
    \begin{subfigure}[b]{0.45\linewidth}
        \centering
        \includegraphics[width=\linewidth]{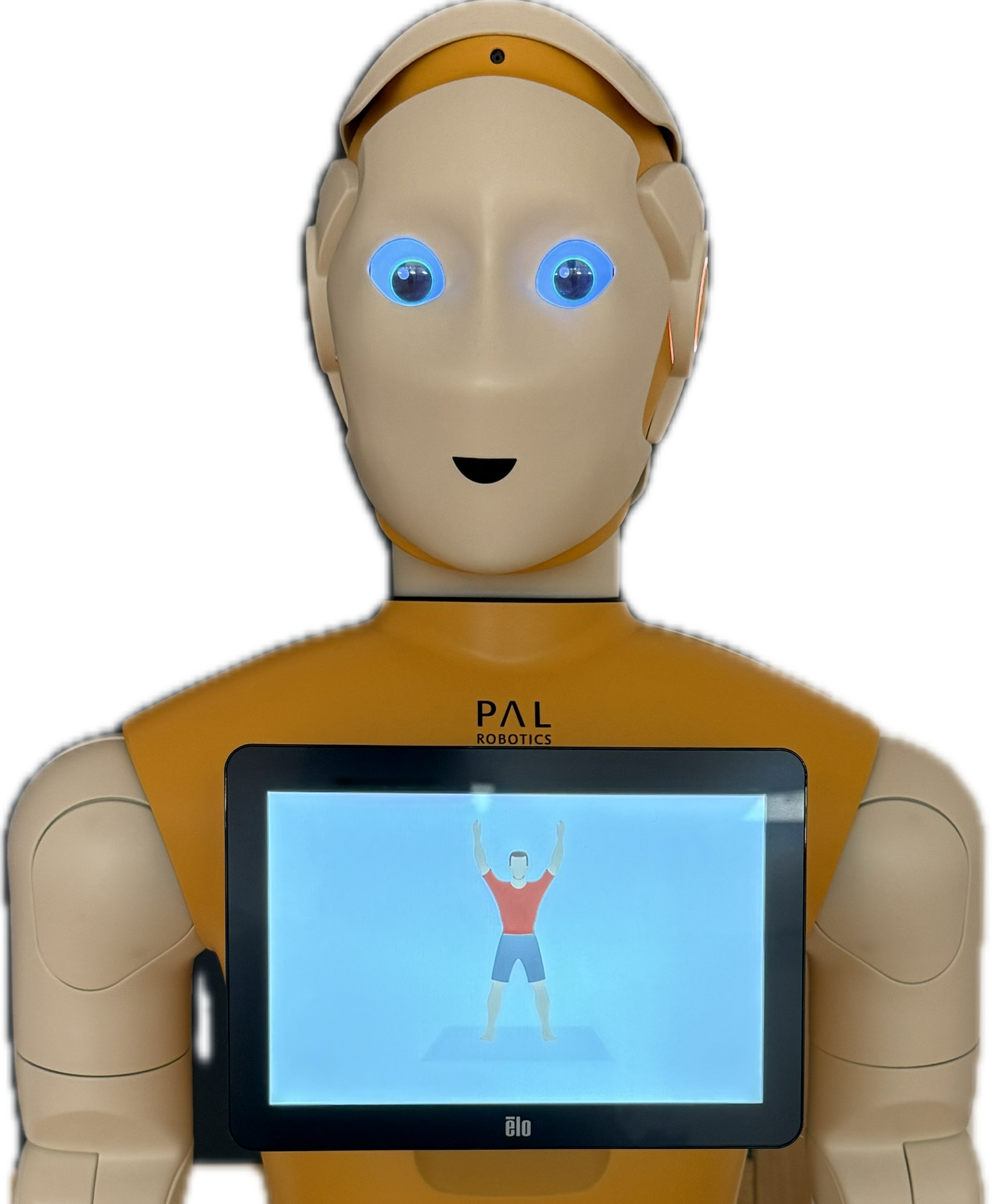}
        \caption{Show exercise}
        \label{fig:exercise}
    \end{subfigure}
    \hfill
    \begin{subfigure}[b]{0.45\linewidth}
        \centering
        \includegraphics[width=\linewidth]{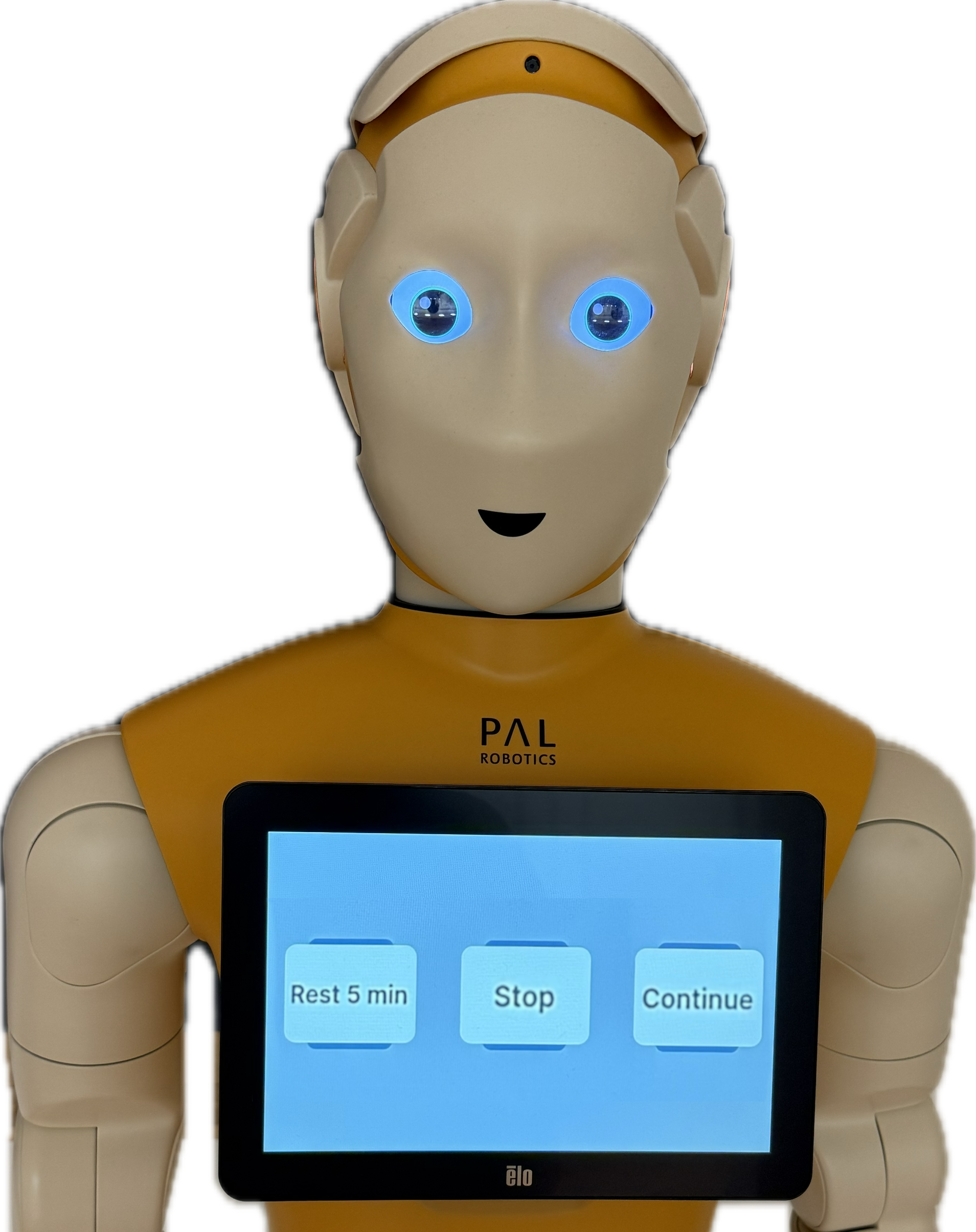}
        \caption{Get input from the user}
        \label{fig:get-input}
    \end{subfigure}
    
    \caption{ARI execution}
    \label{fig:ari}
\end{figure}

We tested the prototype using the representative SLEEC rule (Listing~\ref{lst:sleec-running}), showing that the robot successfully executes the correct obligation based on contextual conditions.
Figure~\ref{fig:ari} illustrates two snapshots from the proof-of-concept development. In Figure~\ref{fig:exercise}, ARI displays the exercise to be performed, guiding the user through the training session in accordance with its main mission. Figure~\ref{fig:get-input} shows ARI asking the user for input on how to proceed (e.g., whether to rest, stop, or continue), as specified by the hedge clause \circnum{2} in Listing~\ref{lst:sleec-running}. The complete demonstration video is available in the replication package~\ReplicationPkg.

Overall, this excerpt of the implementation confirmed the technical applicability of \appNoSpace.
This proof-of-concept thus provides a concrete illustration of how ethical reasoning can be integrated into robotic behavior through \appNoSpace.

\subsection{Threats to Validity}
\label{sec:threats}

We identified the main threats to validity and mitigation strategies according to~\cite{wohlin2012experimentation}, as described in the following.

\smallskip
\noindent\textbf{Internal validity} --
Regarding the \textit{internal validity}, we observed that uncontrollable conditions led to a few cases of high overheads in the \textit{Enforcer} component in the first test cases run during each experiment, as demonstrated by the skewness values. This is likely due to the ramp-up phase of the ASMETA Server (likely caused by the instantiation of the Java objects), the varying system load, and to the operating system's job priority scheduling. This threat was mitigated by running multiple test cases (750 in the scope of EQ2, 5500 across the 110 test cases in the scope of EQ3) in a controlled Docker-based environment, and by monitoring the network-induced delay.

\smallskip
\noindent\textbf{External validity} --
As a threat to the \textit{external validity}, the experiments conducted in a controlled environment using a single deployment strategy may limit the generalizability of the results.
To mitigate this, we designed the \app architecture as a domain-agnostic enforcement layer, while its scalability was evaluated using synthetically generated SLEEC models significantly larger than those expected in realistic deployments.
Moreover, concerning the experiment running, the Test runner was deployed on a Turtlebot 4, representing a ``worst-case'' deployment scenario. However, we also ran 250 test cases on a mid-2020 MacBook Pro equipped with an Intel Core i7 CPU, deploying all components in Docker containers, and still observed limited overhead (0.113 ms on average). Future work will include experiments deploying the Enforcement Subsystem on network edge nodes to simulate further scenarios, and by testing it over heterogeneous SLEEC rulesets.

\smallskip
\noindent\textbf{Construct validity} --
A potential threat for the \textit{construct validity} concerns the potential incompleteness or incorrectness of the elicited SLEEC ruleset for the reference scenario. The defined ruleset may fail to capture all ethically relevant situations of the scenario, potentially allowing the system to entering the ethics-disregarding region (as discussed in \Cref{sec:framework}). We mitigated this threat by (i) following the iterative elicitation process defined in~\cite{Townsend2022sleec}, and (ii) by supporting this process by interviewing domain experts for rule validation. Concerning the validation, subjective opinions reported by the interviewed experts could potentially affect the evaluation. We mitigated this threat by following the qualitative survey empirical standard~\cite{acmsigsoft} and by iteratively interviewing domain experts selected through convenience sampling~\cite{stratton2021population}.

\smallskip
\noindent\textbf{Conclusion validity} --
A threat to the \textit{conclusion validity} concerns the statistical analysis of the scalability results in the scope of EQ3. The regression models used to characterize the overhead may be influenced by the number of generated models, their characteristics (i.e., number of rules and clauses) and by the presence of outliers in the measurements. We mitigated this threat by running 5500 test cases across 110 synthetically generated SLEEC models, reporting multiple goodness-of-fit metrics, and comparing polynomial, linear, and exponential regression models.

\section{Discussion}
\label{sec:discussion}

\noindent\textbf{Runtime planning} --
As described in Section~\ref{sec:architecture}, the Enforcer implements both the analyze and plan phases of the MAPE-K reference model, specifically returning the obligation to be enforced. It is then the responsibility of the Executor to translate this into the sequence of tasks that the system must execute. For autonomous systems, we assume that they have their own planners to dynamically generate execution plans based on their capabilities and functionalities, as for the robots we used for the evaluation. As future work, we plan to extend the approach to enable \emph{interplay between predefined obligations and runtime planning}. 

\smallskip
\noindent\textbf{Temporal constraints on obligations} --
Although at the model level it is possible to specify \emph{time constraints on obligations}, the current ARI-based prototype does not yet implement them. As future work, we plan to integrate a task executor that leverages task waiting and timeout mechanisms available within the ROS-based robotic platform for multi-threading.

\smallskip
\noindent\textbf{\app vs. rule-based approaches} --
Our approach should not be interpreted as a rule-based adaptation framework in which rules constitute the primary mechanism driving system adaptation. Instead, SLEEC rules operate as an ethical enforcement layer that constrains the system’s runtime behavior, ensuring compliance with explicitly specified ethical requirements.
In this sense, \app
does not prescribe nor replace existing adaptation strategies. The underlying autonomous system may adopt any adaptation mechanism, including learning-based, optimization-based, or other advanced approaches. As discussed in \Cref{sec:framework}, our contribution focuses specifically on the ethical-awareness dimension.
This separation between adaptation logic and ethical enforcement promotes modularity and extensibility, allowing ethical requirements to evolve independently from the system’s adaptation mechanisms. 

\smallskip
\noindent\textbf{User-centered ethics enforcement} --
An important implication of \app is the ability to adapt ethical rules independently from the system’s core behavior. Following \cite{Townsend2022sleec}, SLEEC rules are defined based on the system’s capabilities and the application domain, allowing for their evolution if needed, without requiring modifications to the system’s logic. However, SLEEC rules can also be updated to adapt to changes in the operational context (e.g., different environments or scenarios), or to the values and preferences of individual users or groups.
Leveraging \app\ supports flexible integration of ethical considerations, allowing both contextual adaptation and user-specific customization if needed, while avoiding hard-coding ethical decisions into the system behavior.
Additionally, while our current approach directly returns the obligation to be enforced to bring the system back to the ethics-respectful state space, it could be extended to enable the system and the user to \emph{negotiate} either the obligation itself or the sequence of tasks to be executed to enforce it. This extension would provide an additional degree of flexibility.

\section{Related work}
\label{sec:related}

In their vision paper, Boltz et al.~\cite{BoltzYILLZ24} examine how self-adaptive socio-technical systems can be designed to empower humans while respecting diverse needs, values, and ethics. They emphasize the importance of human empowerment and ethical balance at individual, community, and societal levels. Specifically, the authors propose a high-level architectural framework to guide interactions and preserve human values.

Ethical oversight and well-designed regulatory structures are crucial for creating autonomous systems that act responsibly, maintain transparency, and reflect human values~\cite{Dignum25}. With this aim, Townsend et al.~\cite{Townsend2022sleec} proposed a methodology to elicit SLEEC rules as ethical requirements for autonomous systems. 
Since their introduction, multiple researchers have focused on formalizing high-level normative principles into consistent SLEEC rulesets, supporting automated ethical decision-making. Troquard et al.~\cite{Troquard2024aaai-sleec} showed how natural-language SLEEC rules can be converted into classical logic to enable automated normative reasoning. 
Mirani et al.~\cite{MiraniRT24} propose encoding SLEEC rules in Datalog, a declarative logic programming language, to enable scalable computation of obligations.
Yaman et al. introduced the SLEEC-TK toolkit~\cite{GetirYaman2024sleec-tk,GetirYaman2025}, which provides a DSL for specifying SLEEC rules and tools for validating rulesets and verifying design model conformance in tock-CSP~\cite{Miyazawa2019}. Complementarily, Feng et al.~\cite{FengMYSBAMTBCCC24} propose an alternative approach using Large Language Models (LLMs) to enhance automated reasoning techniques to better elicit, analyze, and ensure consistency of normative requirements. 
Kolyakov et al.~\cite{kolyakov2025legos} introduced LEGOS-SLEEC, a tool that assists interdisciplinary stakeholders in defining normative requirements as SLEEC rules and in checking and debugging their consistency.
However, all these approaches remain limited to the elicitation, validation, and verification of SLEEC rules, without operationalizing them as we do in this paper, namely, bringing them into the design, implementation, and execution within autonomous systems, enabling them to perform ethical decision-making.

Many approaches emphasize the importance of incorporating human values into software development. The survey in~\cite{shahin2022operationalizing} notes that while current methods facilitate values operationalization during early stages (requirements and design), later stages, such as implementation and testing, remain underexplored.
Bennaceur et al.~\cite{BennaceurHNZ23} proposed \emph{values@runtime}, an adaptive MAPE-K framework that helps users align decisions with their personal values, detect mismatches, receive recommendations, and reflect on behavior, illustrated via a shopping basket tool.
Complementarily, our approach focuses on adapting the system’s behavior to align with ethical principles, rather than recommending alternative actions to users. Moreover, given the flexibility and runtime nature of our approach, it can be extended to account for users’ values.
\section{Conclusion}\label{sec:conclusion}

This paper addresses the challenge of operationalizing ethics in autonomous systems through runtime enforcement. It instantiates \app, an ethics assurance process that spans from the elicitation of ethical requirements (formalized as SLEEC rules) to their enactment during system operation. Leveraging the Abstract State Machine formalism and its supporting toolset, the approach enables the dynamic evaluation, adaptation, and enforcement of ethical behavior.
We validated the proposed framework through robotic experiments conducted both in a simulated environment and on a real humanoid robot. The results show that delegating the enforcement of ethical rules to an enforcement subsystem ensures compliance with ethical principles at negligible runtime cost.

Although the case study focuses on rehabilitation support, the developed assistive robot can be extended to other domains as a multi-service platform. Such extensions would entail additional SLEEC rules to regulate ethical behavior under new operational contexts.

\section*{Data Availability Statement}
A replication package containing the SLEEC requirements, SLEEC model in ASMETA, all software artifacts, and data sets used to evaluate our approach is available online at~\ReplicationPkg.

\section*{Acknowledgments}
This work has been partially funded by 
(a) the MUR (Italy) Department of Excellence 2023 - 2027, 
(b) the PRIN project P2022RSW5W - RoboChor: Robot Choreography, 
(c) the PRIN project 2022JKA4SL - HALO: etHical-aware AdjustabLe autOnomous systems,
(d) the Helmholtz Association (HGF) with the KiKIT project,
(e) the HGF Grant 46.23 (Engineering Secure Systems), and
(f) the PRIN 2022 PNRR project SAFEST: truSt Assurance of digital twins For mEdical cyber-phySical sysTems  (G53D23002770006 and F53D23004230006).

\bibliographystyle{ieeetr}
\bibliography{references}

\vfill

\end{document}